\documentclass[aps,prapplied,twocolumn,10pt,groupedaddress,noeprint]{revtex4-2}

\usepackage{graphicx}
\usepackage{amsfonts}
\usepackage{amsmath}
\usepackage{amssymb}
\usepackage{mathtools}
\usepackage{siunitx}
\usepackage{enumerate}
\usepackage{cases}
\usepackage{bm}
\usepackage{empheq}
\usepackage{multirow}
\usepackage{physics}
\usepackage[pdfusetitle,colorlinks,allcolors=blue,pdftex,
pdfauthor={Matteo Mariantoni},
pdftitle={Fluctuation Spectroscopy of Two-Level Systems in Superconducting Resonators},
pdfkeywords={Quantum Computing, Superconducting Resonators, Lumped Elements, Superconducting Qubits, Amorphous Dielectric Materials, Two-Level Systems, Standard Tunneling Model, Long-Time Stochastic Fluctuations, Two-Tone Measurements, Fluctuation Spectroscopy, 1/f Noise, Generalized Tunneling Model}]{hyperref}

\newcommand{\round}[1]{\ensuremath{\lfloor#1\rceil}}

\begin{document}

\title{Fluctuation Spectroscopy of Two-Level Systems in Superconducting Resonators}

\author{J. H.~B\'{e}janin}
\thanks{These two authors contributed equally to this work.}
\affiliation{Institute for Quantum Computing, University of Waterloo, 200 University Avenue West, Waterloo, Ontario N2L 3G1, Canada}
\affiliation{Department of Physics and Astronomy, University of Waterloo, 200 University Avenue West, Waterloo, Ontario N2L 3G1, Canada}

\author{Y.~Ayadi}
\thanks{These two authors contributed equally to this work.}
\affiliation{Institute for Quantum Computing, University of Waterloo, 200 University Avenue West, Waterloo, Ontario N2L 3G1, Canada}
\affiliation{Department of Physics and Astronomy, University of Waterloo, 200 University Avenue West, Waterloo, Ontario N2L 3G1, Canada}

\author{X.~Xu}
\affiliation{Institute for Quantum Computing, University of Waterloo, 200 University Avenue West, Waterloo, Ontario N2L 3G1, Canada}
\affiliation{Department of Physics and Astronomy, University of Waterloo, 200 University Avenue West, Waterloo, Ontario N2L 3G1, Canada}

\author{C.~Zhu}
\affiliation{Institute for Quantum Computing, University of Waterloo, 200 University Avenue West, Waterloo, Ontario N2L 3G1, Canada}
\affiliation{Department of Physics and Astronomy, University of Waterloo, 200 University Avenue West, Waterloo, Ontario N2L 3G1, Canada}

\author{H. R.~Mohebbi}
\affiliation{Institute for Quantum Computing, University of Waterloo, 200 University Avenue West, Waterloo, Ontario N2L 3G1, Canada}
\affiliation{High Q Technologies LP, 485 Wes Graham Way, Waterloo, Ontario N2L 0A7, Canada}

\author{M.~Mariantoni}
\email[Corresponding author: ]{matteo.mariantoni@uwaterloo.ca}
\affiliation{Institute for Quantum Computing, University of Waterloo, 200 University Avenue West, Waterloo, Ontario N2L 3G1, Canada}
\affiliation{Department of Physics and Astronomy, University of Waterloo, 200 University Avenue West, Waterloo, Ontario N2L 3G1, Canada}

\date{\today}

\begin{abstract}
Superconducting quantum computing is experiencing a tremendous growth. Although major milestones have already been achieved, useful quantum-computing applications are hindered by a variety of decoherence phenomena. Decoherence due to two-level systems~(TLSs) hosted by amorphous dielectric materials is ubiquitous in planar superconducting devices. We use high-quality quasilumped element resonators as quantum sensors to investigate TLS-induced loss and noise. We perform two-tone experiments with a probe and pump electric field; the pump is applied at different power levels and detunings. We measure and analyze time series of the quality factor and resonance frequency for very long time periods, up to~\SI{1000}{\hour}. We additionally carry out simulations based on the TLS interacting model in presence of a pump field. We find that loss and noise are reduced at medium and high power, matching the simulations, but not at low power.
\end{abstract}


\maketitle

\section{INTRODUCTION}
	\label{Sec:INTRODUCTION}

Superconducting quantum circuits~\cite{Rasmussen:2021} are among the leading systems in the race to develop a quantum computer; other relevant examples include trapped atomic ions and neutral atoms~\cite{Alexeev:2021}. Although exciting and potentially disruptive, applications such as quantum simulations require a daunting amount of resources and complex infrastructures~\cite{Altman:2021}. In contrast, an immediate application of superconducting devices is to use a single quantum circuit as a sensor for loss and noise processes in dielectric and superconducting materials~\cite{Murray:2021}. Understanding such dissipative phenomena, which lead to quantum decoherence, is of paramount importance to further scale up quantum computers.

Superconducting devices based, for instance, on aluminum~(Al) or niobium thin films deposited on silicon~(Si) or sapphire substrates can be readily fabricated as integrated circuits; important examples include the Xmon transmon qubit~\cite{Barends:2013} and planar resonators~\cite{McRae:2020}. On-chip integration, however, comes at a price: Almost any interface separating different fabrication layers is characterized by unwanted amorphous dielectric materials, such as native silicon oxide~\cite{Earnest:2018}. In fact, one of the main sources of decoherence is attributable to two-level systems~(TLSs) embedded in oxide layers~\cite{Phillips:1987,Mueller:2019}. Interestingly, deleterious TLS effects are also impactful in other quantum computing platforms, such as trapped ions~\cite{Noel:2019} and optical cavities~\cite{Kong:2021}. Therefore, these implementations can also benefit from the research presented here.

Defects in amorphous dielectric materials, such as trapped charges, dangling bonds, tunneling atoms, or collective motion of molecules, can be modeled as TLSs with energy~$E$. We distinguish between quantum-TLSs~(Q-TLSs), for which~$E > k_{\text{B}} T$, and thermal-TLSs~(T-TLSs), for which~$E < k_{\text{B}} T$; the typical operating temperature of our devices is~$T \sim \SI{50}{\milli\kelvin}$ and, thus, the energy threshold between Q- and T-TLSs is~$E/h \sim \SI{1}{\giga\hertz}$~\cite{Bejanin:2021}.

Interactions with TLSs reduce the coherence properties of superconducting quantum circuits, resulting in shorter energy relaxation times~$T_1$ for qubits and lower internal quality factors~$Q^{\text{r}}_{\text{int}}$ for resonators. It is well known that driving resonators electrically saturates TLSs, leading to a higher~$Q^{\text{r}}_{\text{int}}$~\cite{Mueller:2019}; this can also be accomplished via an off-resonant drive~\cite{Sage:2011,Kirsh:2017,Capelle:2020}. The effect of TLSs is not limited to loss mechanisms but also leads to noise (i.e., time fluctuations) in~$Q^{\text{r}}_{\text{int}}$ and in the resonator resonance frequency~$f_{\text{r}}$~\cite{Neill:2013,Burnett:2014,Moeed:2019}; qubits exhibit an equivalent behavior~\cite{Paik:2011,Klimov:2018,Burnett:2019,Schloer:2019,Bejanin:2021,Carroll:2021}. Our main motivation is to gain a deeper insight into the physics of TLS-induced time fluctuations for much longer time periods and a wider range of parameters compared to these previous works.

In this paper, we conduct a \emph{fluctuation spectroscopy} experiment with quasilumped element resonators based on planar capacitors and inductors. The resonators are characterized by~$Q^{\text{r}}_{\text{int}} \sim 350000$ when excited at low power. We perform a two-tone experiment by driving the resonators with a probe and pump electric field. The probe is used to measure~$f_{\text{r}}$ and $Q^{\text{r}}_{\text{int}}$ and is always set to low power. The pump is used to excite Q-TLSs and can take different power values~$P_{\text{p}}$ and resonator-pump detunings~$\Delta_{\text{r,p}}$. We investigate time fluctuations in the time series~$f_{\text{r}} ( t )$ and $Q^{\text{r}}_{\text{int}} ( t )$ for two resonators over a total observation time of~\SI{480}{\hour}; this time comprises four periods that are spread over one and a half months. Such a long measurement time allows us to accurately estimate the power spectral densities~(PSDs) associated with the time series to frequencies as low as~$f = \SI{5.8e-6}{\hertz}$.

The present work represents a significant departure from our previous study of Ref.~\cite{Bejanin:2021}. There, we have investigated the~$T_1$ (equivalent to~$Q^{\text{r}}_{\text{int}}$) spectrotemporal charts of an Xmon transmon qubit and compared them to detailed simulations based on the generalized tunneling model~(GTM). Here, we add a pump, measure~$f_{\text{r}}$ (which we could not measure with an Xmon qubit), and extend the time series to a much longer time. We compare our experimental findings to GTM simulations, accounting for these more general conditions.

We find that the simulations match the experimental time series and PSDs for both~$f_{\text{r}}$ and $Q^{\text{r}}_{\text{int}}$ when the pump is off. At low pump power, the loss and time fluctuations are not reduced compared to when the pump is off. In fact, in certain cases, we even observe a small increase in loss and fluctuations. This result slightly deviates from our simulations, which display a monotonic decrease in loss and fluctuations with increasing pump power. At medium and high power, as expected, loss and fluctuations are reduced both in experiments and simulations. Interestingly, for the same value of~$P_{\text{p}}$ but different~$\Delta_{\text{r,p}}$, we measure dissimilar time series and, often, even different noise levels. These asymmetries are not reproduced by the simulations. Finally, it is worth noting that we measure~$\sim 1/f$ noise down to microhertz frequencies.

Recent works have investigated time fluctuations in presence of a pump~\cite{Burnett:2014,DeGraaf:2018,Niepce:2021}. However, these authors only use a resonant pump and carry out measurements over short time periods, on the order of~\SI{3}{\hour}. Additionally, they only study the low and medium power regimes but not the pump off and high power regimes. Finally, they do not attempt to match their experimental results to any simulations.

The paper is organized as follows. In Sec.~\ref{Sec:THEORY}, we review the models governing resonator--Q-TLS and Q-TLS--T-TLS interactions. In Sec.~\ref{Sec:METHODS}, we explain the methods used to perform experiments, simulations, and spectral analysis. In Sec.~\ref{Sec:RESULTS}, we present our results. In Sec.~\ref{Sec:DISCUSSION}, we suggest a possible explanation of some of our findings and compare them to recent related works. Finally, in Sec.~\ref{Sec:CONCLUSIONS}, we summarize our results and propose future research directions.

\section{THEORY}
	\label{Sec:THEORY}

Planar superconducting devices characterized by oxide layers offer a versatile environment for the study of TLSs. Each TLS can be modeled as a double-well potential with tunneling and asymmetry energies~$\Delta_0$ and $\Delta$, respectively. The diagonalized TLS Hamiltonian reads as~$\widehat{H}_{\text{TLS}} = E \, \hat{\sigma}_z / 2$, where~$\hat{\sigma}_z$ is the Pauli matrix in the energy basis and $E = \sqrt{\Delta^2_0 + \Delta^2}$ is the TLS energy.

Q-TLSs interact semiresonantly with quantum microwave resonators and, at the same time, with far-off-resonant T-TLSs. The first interaction induces decoherence as well as a frequency shift in the resonator (Sec.~\ref{Subsec:Driven-dissipative:resonator--Q-TLS:interaction}). The second interaction leads to stochastic time fluctuations and, thus, a dependence on time~$t$ in the transition frequency of a Q-TLS, $f_{\text{Q-TLS}} ( t )$ (Sec.~\ref{Subsec:Resonator:stochastic:fluctuations}).

\subsection{Driven-dissipative resonator--Q-TLS interaction}
	\label{Subsec:Driven-dissipative:resonator--Q-TLS:interaction}

In general, an ensemble of Q-TLSs coupled to one resonator can be represented by a Tavis-Cummings Hamiltonian. This model can be further simplified by assuming that the Q-TLSs interact with the resonator \emph{nonsimultaneously}, thereby leading to a set of independent Jaynes-Cummings Hamiltonians. Different regimes of the Q-TLS--resonator system can be investigated by means of a classical driving field, or pump.

The pump is a monochromatic sinusoidal signal with voltage amplitude~$V_{\text{p}}$ and current amplitude~$I_{\text{p}}$, frequency~$f_{\text{p}}$, and time-averaged power~$P_{\text{p}} = V_{\text{p}} I_{\text{p}} / 2$. In our physical system, we pump indirectly the Q-TLSs by driving the resonator (sympathetic driving). We can explore the on-resonance and off-resonance regimes by tuning~$f_{\text{p}}$ as well as the high-power and low-power regimes by varying~$P_{\text{p}}$. The pump Hamiltonian reads as
\begin{equation}
\widehat{H}_{\text{p}} = h J \! \cos\!\left( 2 \pi f_{\text{p}} t \right) \! \left( \hat{a}^{\dagger} + \hat{a} \right) ,
	\label{eq:Hp}
\end{equation}
where~$J = \sqrt{\Phi_{\text{p}} \Gamma^{\text{r}}_{\text{ext}} / 2}$ is the coupling coefficient between a quasi--one-dimensional transmission line and the resonator (see Sec.~\ref{Subsec:METHODS:Resonator:measurement} for device details~\footnote{The~$1 / \sqrt{2}$ factor in~$J$ is due to the hanger-type coupling configuration~\cite{McRae:2020} of our resonators.}), and $\hat{a}^{\dagger}$ and $\hat{a}$ are the usual bosonic annihilation and creation operators of the resonator mode~\footnote{In our experiments, the resonators are characterized by a single mode because they are made of quasilumped elements.}; $\Phi_{\text{p}} = P_{\text{p}} / h f_{\text{p}}$ is the transmission line photon flux, and $\Gamma^{\text{r}}_{\text{ext}}$ is the engineered, or \emph{external}, energy relaxation rate of the resonator.

In addition to the engineered losses described by~$\Gamma^{\text{r}}_{\text{ext}}$, the resonator suffers from internal dissipation processes. These processes, which are mainly due to conductive, radiative, and dielectric losses, are described by the \emph{internal} decoherence rate~$\Gamma^{\text{r}}_{\text{int}}$. Among all types of dielectric losses, those strongly depending on~$P_{\text{p}}$ are predominantly caused by Q-TLSs. These losses are associated with an energy relaxation rate~$\Gamma^{\text{r,Q-TLS}}_{\text{int}}$, whereas all other losses are accounted for by~$\widetilde{\Gamma}^{\text{r}}_{\text{int}}$. In the case of high quality resonators, such as those studied in this work, the resonator pure dephasing rate is negligible and, thus, $\Gamma^{\text{r}}_{\text{int}}$ is assumed to be an energy relaxation rate~\cite{Wang:2008}.

The rate~$\Gamma^{\text{r,Q-TLS}}_{\text{int}}$ is given by the sum of partial decoherence rates, where each rate~$\kappa^{\text{r,Q-TLS}}_{\text{int}}$ is due to the interaction between one Q-TLS and the resonator. This interaction additionally leads to a shift~$\delta \! f_{\text{r}}$ of the resonance frequency. Both~$\kappa^{\text{r,Q-TLS}}_{\text{int}}$ and $\delta \! f_{\text{r}}$ can be calculated by solving for the time evolution of a driven-dissipative Jaynes-Cummings interaction. In a reference frame rotating with~$f_{\text{p}}$ and after a rotating wave approximation, the driven Jaynes-Cummings Hamiltonian reads~as
\begin{eqnarray}
\widehat{\!\widetilde{H}}_{\text{r,Q-TLS}}
	& = & h \dfrac{\Delta_{\text{Q-TLS,p}}}{2} \hat{\sigma}_z + h \Delta_{\text{r,p}} \, \hat{a}^{\dagger} \hat{a}
	\nonumber\\
	&   & + h g \left( \hat{\sigma}^+ \hat{a} + \hat{\sigma}^- \hat{a}^{\dagger} \right) + h J \left( \hat{a}^{\dagger} + \hat{a} \right) ,
	\label{eq:Htilde:r:Q-TLS}
\end{eqnarray}
where~$\Delta_{\text{Q-TLS,p}} = f_{\text{Q-TLS}} ( t ) - f_{\text{p}}$, $\Delta_{\text{r,p}} = \tilde{f}_{\text{r}} - f_{\text{p}}$, $g$ is the coupling strength of the Q-TLS--resonator interaction, and $\hat{\sigma}^{\mp}$ are the Q-TLS lowering and raising operators in the energy basis; $\tilde{f}_{\text{r}}$ is the unperturbed resonance frequency of the resonator.

The dissipative time evolution is obtained from the Lindbladian
\begin{equation}
\dfrac{d \hat{\rho}}{dt} = - \dfrac{i}{\hbar} \left[ \widehat{\!\widetilde{H}}_{\text{r,Q-TLS}} , \hat{\rho} \right] + \sum_{\hat{L}} \left( \hat{L} \hat{\rho} \hat{L}^{\dagger} - \dfrac{1}{2} \left\{\hat{L}^{\dagger} \hat{L} , \hat{\rho}\right\} \right) ,
	\label{eq:rhodot}
\end{equation}
where~$\hat{\rho}( t )$ is the time dependent density matrix, $\widehat{\!\widetilde{H}}_{\text{r,Q-TLS}}$ is given by Eq.~(\ref{eq:Htilde:r:Q-TLS}), and $\hat{L}$ and $\hat{L}^{\dagger}$ are Lindblad operators with~$\hat{L} \in \{ \hat{L}_{\text{r}} , \hat{L}^-_{\text{Q-TLS}} \}$. For simplicity, we disregard the pure dephasing rate of the Q-TLS, $\Gamma^{\text{Q-TLS}}_{\phi} = 0$. Under these conditions, the Lindblad operators are~$\hat{L}_{\text{r}} = \sqrt{\Gamma^{\text{r}}_{\text{ext}}} \, \hat{a}$ and $\hat{L}^-_{\text{Q-TLS}} = \sqrt{\Gamma^{\text{Q-TLS}}_1} \, \hat{\sigma}^-$, where~$\Gamma^{\text{Q-TLS}}_1$ is the energy relaxation rate of the Q-TLS for~$T \rightarrow 0^+$; $\tilde{f}_{\text{Q-TLS}} = E_{\text{Q-TLS}} / h$ is the unperturbed Q-TLS transition frequency.

After solving the quantized Maxwell-Bloch equations in the stationary regime (see App.~\ref{App:Q-TLS:PARTIAL:CONTRIBUTIONS}), we find the~$2$-tuple~$( \kappa^{\text{r,Q-TLS}}_{\text{int}} , \delta \! f_{\text{r}} )$. Such a tuple can be associated with the~$k$th Q-TLS of an ensemble of Q-TLSs, $( \kappa^{\text{r,Q-TLS}}_{\text{int}} \rightarrow \kappa^{\text{r} , k}_{\text{int}} , \delta \! f_{\text{r}} \rightarrow \delta \! f_{\text{r} , k} )$. Each Q-TLS is characterized by a coupling strength~$g_k$, energy relaxation rate~$\Gamma^k_1$, and frequency~$f_k ( t )$.

The collective effect of all Q-TLSs gives
\begin{subequations}
	\begin{empheq}[]{align}
		\Gamma^{\text{r}}_{\text{int}} & = \widetilde{\Gamma}^{\text{r}}_{\text{int}} + \Gamma^{\text{r,Q-TLS}}_{\text{int}} = \widetilde{\Gamma}^{\text{r}}_{\text{int}} + \sum_k \kappa^{\text{r} , k}_{\text{int}}
	\label{subeq:Gamma:r:int}
	\end{empheq}
\end{subequations}
and results in the instantaneous resonance frequency
\setcounter{equation}{\value{equation}-1}
\begin{subequations}
\setcounter{equation}{1}
	\begin{empheq}[]{align}
		f_{\text{r}} & = \tilde{f}_{\text{r}} + \Delta f_{\text{r}} = \tilde{f}_{\text{r}} + \sum_k \delta \! f_{\text{r} , k} .
	\label{subeq:f:r}
	\end{empheq}
\end{subequations} 

Given~$\Gamma^{\text{r}}_{\text{tot}} = \Gamma^{\text{r}}_{\text{ext}} + \Gamma^{\text{r}}_{\text{int}}$, the intraresonator mean photon number~$\langle n \rangle$ can be obtained by solving the quantized Maxwell-Bloch equations for~$g = 0$:
\begin{equation}
\langle n \rangle = \frac{2 \Phi_{\text{p}} \Gamma^{\text{r}}_{\text{ext}}}{16 \pi^2 \Delta^2_{\text{r,p}} \! + \! \left( \Gamma^{\text{r}}_{\text{tot}} \right)^2} .
	\label{eq:mean:n}
\end{equation}

\subsection{Resonator stochastic fluctuations}
	\label{Subsec:Resonator:stochastic:fluctuations}

In Sec.~\ref{Subsec:Driven-dissipative:resonator--Q-TLS:interaction}, we describe the interaction between a resonator and an ensemble of Q-TLSs. Considering the model in our previous work of Ref.~\cite{Bejanin:2021}, we further assume that each Q-TLS strongly interacts with one or more T-TLSs according to the GTM. The interaction with one T-TLS shifts~$\tilde{f}_{\text{Q-TLS}}$ by~$\delta \! f^{\mp}$, where the sign depends on the state~$\ket{\mp}$ of the T-TLS. This state switches in time due to thermal activation and can be modeled by a random telegraph signal~(RTS) with switching rate~$\gamma$; therefore, the frequency shift varies in time also as an~RTS, $\delta \! f^{\mp} ( t )$.

The~$\ell_k$th T-TLS of all the T-TLSs coupled to the~$k$th Q-TLS is characterized by a~$2$-tuple of fundamental parameters~$( \delta \! f^{\mp} \rightarrow \delta \! f^{\mp}_{\ell_k} , \gamma \rightarrow \gamma_{\ell_k} )$. The effective frequency of the~$k$th Q-TLS due to the interaction with the T-TLSs can be written as a time series,
\begin{equation}
f_k ( t ) = \tilde{f}_k + \sum_{\ell_k} \delta \! f^{\mp}_{\ell_k} ( t ) ,
	\label{eq:fkt}
\end{equation}
where~$\tilde{f}_k$ is the unperturbed transition frequency of the~$k$th Q-TLS. The stochastic fluctuations inherent to Eq.~(\ref{eq:fkt}) are present in Eq.~(\ref{eq:sigma:z0}) as well as in Eqs.~(\ref{subeq:kappa:rQ-TLS:int}) and (\ref{subeq:2pi:delta:f:r}); thus, these fluctuations propagate to the sums in Eqs.~(\ref{subeq:Gamma:r:int}) and (\ref{subeq:f:r}) leading to the time series~$\Gamma^{\text{r}}_{\text{int}} ( t )$ and $f_{\text{r}} ( t )$. The former replaces Eq.~(9) in our work of Ref.~\cite{Bejanin:2021} to account for the pump field, whereas the latter describes frequency shifts not presented in that work. These time series constitute the main model for the experiments and simulations presented in this paper.

\section{METHODS}
	\label{Sec:METHODS}

In this section, we describe the methods used to perform the resonator measurement (Sec.~\ref{Subsec:METHODS:Resonator:measurement}), the off-resonant pumping (Sec.~\ref{Subsec:METHODS:Off-resonant:pumping}), the simulations (Sec.~\ref{Subsec:METHODS:Simulations}), and the spectral analysis (Sec.~\ref{Subsec:METHODS:Spectral:analysis}).

\subsection{Resonator measurement}
	\label{Subsec:METHODS:Resonator:measurement}

In this work, we perform experiments on a set of superconducting microwave resonators. The resonators are implemented as quasilumped element planar structures, comprising an inductor with inductance~$L_{\text{r}}$ and a capacitor with capacitance~$C_{\text{r}}$.

We study hanger-type resonators coupled by means of a capacitor with capacitance~$C_{\text{ext}}$ to a coplanar waveguide~(CPW) transmission line~\cite{McRae:2020}---the feed line; this line has a characteristic impedance~$Z_0$. The resonance frequency of each coupled resonator is~$\tilde{f}_{\text{r}} \simeq 1 / 2 \pi \sqrt{L_{\text{r}} \left( C_{\text{r}} + C_{\text{ext}} \right)}$. Six resonators are arranged in a frequency multiplexed design, where each resonator has a different~$\tilde{f}_{\text{r}}$. We measure only the two highest quality resonators, which are labeled~R1 and R2 in Fig.~\ref{Fig:App:Sample:micrographs}. Sample and setup details are given in App.~\ref{App:SAMPLE:AND:SETUP}; sample parameters are reported in Table~\ref{Tab:App:Sample:and:characterization:parameters}.

The measurement of a resonator is conducted by means of a probe field with frequency~$f_{\text{pr}}$ and voltage amplitude~$V_{\text{pr}}$. This field interacts with the resonator according to the Hamiltonian of Eq.~(\ref{eq:Hp}), with~$f_{\text{p}} \rightarrow f_{\text{pr}}$ and $V_{\text{p}} \rightarrow V_{\text{pr}}$.

In the experiments, the probe field is generated by a vector network analyzer~(VNA). We use this field to measure the frequency-dependent transmission coefficient~$S_{21}$ of the resonator by sweeping~$f_{\text{pr}}$ over a narrow bandwidth centered at the resonance frequency. To improve data reliability in the low-transmission region of the Lorentzian resonance curve, we double the number of measurement points in the middle one-third section compared to the two side sections of the curve. This measurement is performed at a probe power~$P_{\text{pr}} = V^2_{\text{pr}} / 2 Z_0$.

In order to extract resonator parameters, we analyze the~$S_{21}$ data using the normalized fitting function
\begin{equation}
\widetilde{S}_{21} = \left( 1 + \frac{Q^{\text{r}}_{\text{int}}}{Q^{\text{r}}_{\text{ext}}} e^{i \phi} \frac{f_{\text{pr}}}{f_{\text{pr}} - i 2 Q^{\text{r}}_{\text{int}} \Delta_{\text{r,pr}}} \right)^{\!\!-1} ,
	\label{eq:S21tilde}
\end{equation}
where~$Q^{\text{r}}_{\text{int}}$ and $Q^{\text{r}}_{\text{ext}}$ are, respectively, the internal and external quality factors of the resonator, $\phi$ is the impedance mismatch angle, and $\Delta_{\text{r,pr}} = f_{\text{r}} - f_{\text{pr}}$. Details on the model and analysis procedure can be found in the works of Refs.~\cite{Megrant:2012,Earnest:2018}. The quality factors are related to the energy relaxation rates by
\begin{displaymath}
\Gamma^{\text{r}}_{\text{int}} = \frac{2 \pi f_{\text{r}}}{Q^{\text{r}}_{\text{int}}} \qand \Gamma^{\text{r}}_{\text{ext}} = \frac{2 \pi f_{\text{r}}}{Q^{\text{r}}_{\text{ext}}} .
\end{displaymath}

We characterize each resonator over a wide range of~$P_{\text{pr}}$ (see Fig.~\ref{Fig:App:S-curve:measurements}). In the high-power regime, the resonator--Q-TLS interaction is largely suppressed due to Q-TLS saturation. This regime allows us to estimate the unperturbed parameters~$\widetilde{\Gamma}^{\text{r}}_{\text{int}}$ and $\tilde{f}_{\text{r}}$ in Eqs.~(\ref{subeq:Gamma:r:int}) and (\ref{subeq:f:r}). We estimate~$\Gamma^{\text{r}}_{\text{ext}}$ also in the high-power regime, although this rate should be power independent. In the low-power regime, energy relaxation is dominated by Q-TLS interactions. Sweeping~$P_{\text{pr}}$ between the two regimes makes it possible to evaluate the total energy relaxation rate of the resonator \emph{due to Q-TLSs} at zero mean photon number (i.e., zero power) and zero temperature, $\Gamma^{\text{r,Q-TLS}}_{\text{int}} ( \langle n \rangle \! = \! 0 , T \! = \! \SI{0}{\kelvin} ) = \Gamma^0_{\text{Q-TLS}}$. Additionally, an approximate value for the critical photon number~$n_{\text{c}}$, $n_{\text{c}} \sim \langle n_{\text{s}} \rangle$ ($n_{\text{s}}$ is defined in App.~\ref{App:Q-TLS:PARTIAL:CONTRIBUTIONS}), can be found. The results of this characterization are given in App.~\ref{App:SAMPLE:AND:SETUP}; characterization parameters are reported in Table~\ref{Tab:App:Sample:and:characterization:parameters}. Notably, our~$Q^{\text{r}}_{\text{int}}$ are on the order of~$350000$ at low power, which is quite high for lumped-element resonators (e.g., significantly higher than in the work of Ref.~\cite{Capelle:2020}).

\subsection{Off-resonant pumping}
	\label{Subsec:METHODS:Off-resonant:pumping}

Off-resonant pumping is implemented by linearly superposing the pump to the probe field. The pump (see Sec.~\ref{Subsec:Driven-dissipative:resonator--Q-TLS:interaction}) allows us to indirectly drive the Q-TLS distribution for a range of values of~$\Delta_{\text{r,p}}$ and $P_{\text{p}}$. The probe is solely used to extract resonator parameters by measuring~$S_{21}$ (see Sec.~\ref{Subsec:METHODS:Resonator:measurement}). For all experiments, we use a fixed value of~$P_{\text{pr}}$ such that, when the pump is off ($P_{\text{p}} = \SI{0}{dBm}$), $\langle n \rangle \sim n_{\text{c}}$.

\begin{figure*}[ht!]
	\centering
	\includegraphics{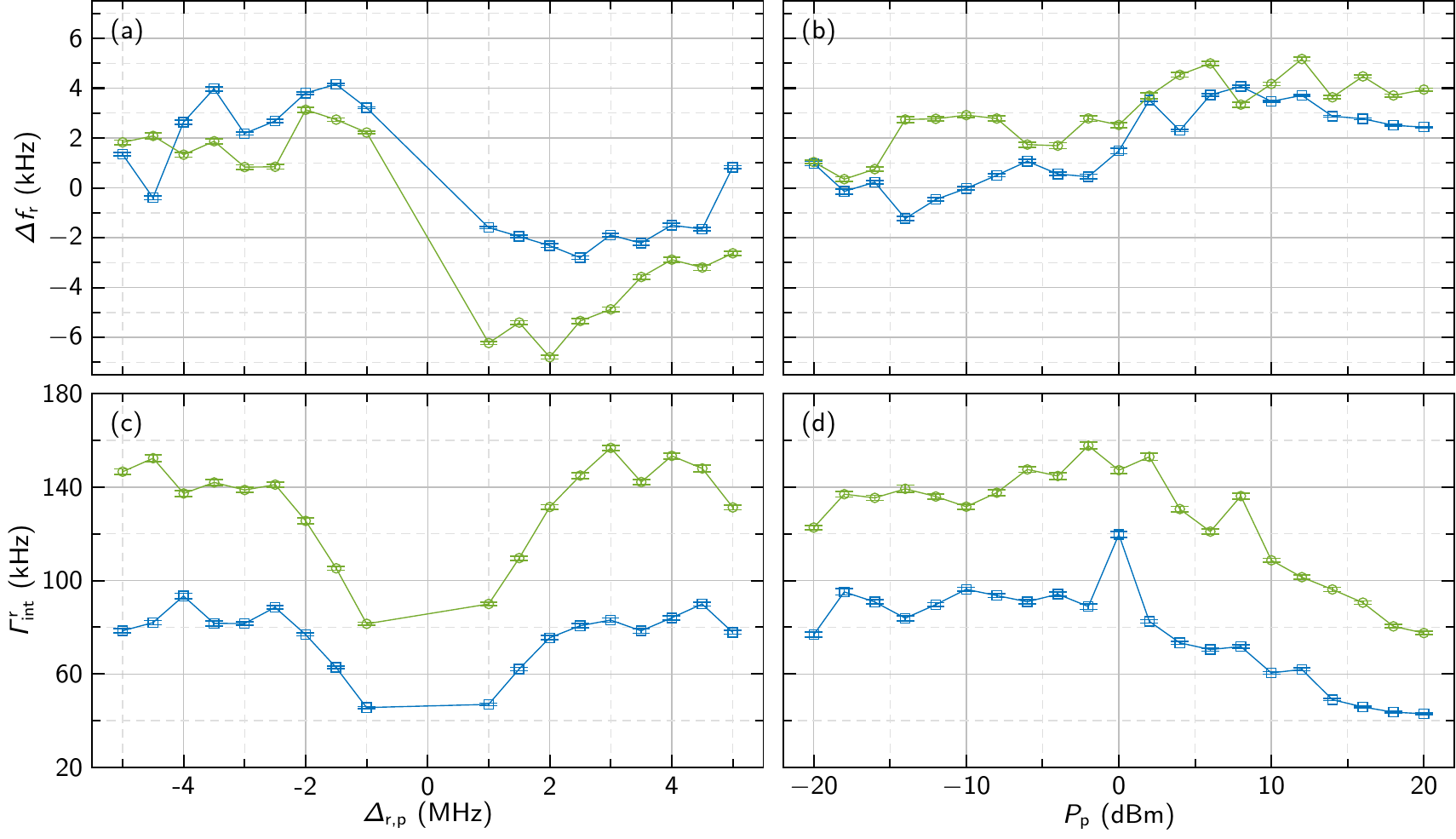}
	\caption{Pumped resonator characterization for~R1 (blue squares) and R2 (green circles). The values are obtained by fitting resonator measurements using Eq.~(\ref{eq:S21tilde}). (a) $\Delta f_{\text{r}}$ vs.~$\Delta_{\text{r,p}}$ at~$P_{\text{p}} = \SI{5}{dBm}$. (b) $\Delta f_{\text{r}}$ vs.~$P_{\text{p}}$ at~$\Delta_{\text{r,p}} = \SI{-2}{\mega\hertz}$. (c) $\Gamma^{\text{r}}_{\text{int}}$ vs.~$\Delta_{\text{r,p}}$ at~$P_{\text{p}} = \SI{5}{dBm}$. (d) $\Gamma^{\text{r}}_{\text{int}}$ vs.~$P_{\text{p}}$ at~$\Delta_{\text{r,p}} = \SI{-2}{\mega\hertz}$. Error bars represent~\SI{95}{\percent} confidence intervals. Lines are guides to the eye. \label{Fig:METHODS:Pumped:resonator:characterization}}
\end{figure*}

Figure~\ref{Fig:METHODS:Pumped:resonator:characterization} shows the characterization of~R1 and R2 in presence of a pump field. Our measurements reproduce the results reported in the works of Refs.~\cite{Sage:2011,Kirsh:2017,Capelle:2020}. As in those works, we also find that the rate~$\Gamma^{\text{r}}_{\text{int}}$ decreases with increasing~$P_{\text{p}}$ and with decreasing~$\abs{\Delta_{\text{r,p}}}$. The frequency~$f_{\text{r}}$ is pulled towards~$f_{\text{p}}$; the frequency shift magnitude~$\abs{\Delta f_{\text{r}}}$ depends on~$P_{\text{p}}$, with the maximum shift occurring at an intermediate value of~$P_{\text{p}}$.

In order to avoid interference effects between the pump and probe fields, $f_{\text{p}}$ must be outside the frequency bandwidth of the~$S_{21}$ resonator measurement. In our measurements, this bandwidth is at most approximately~\SI{0.7}{\mega\hertz} including normalization (i.e., calibration) points; thus, we choose the minimum detuning to be~$\abs{\Delta_{\text{r,p}}} = \SI{1}{\mega\hertz}$. In addition, we select the power of the microwave source used to generate the pump field, $P_{\text{ms}}$, such that~$P_{\text{ms}} \leq \SI{20}{dBm}$ for both resonators to remain below the~$1$-\si{\decibel} compression point of the amplification chain (see Sec.~S3 of the Supplemental Material at~\cite{SM} for more details on compression). Due to this constraint, we cannot reach complete Q-TLS saturation in the high-power regime of Fig.~\ref{Fig:METHODS:Pumped:resonator:characterization}~(b).

We investigate stochastic fluctuations and pumping effects in the time series~$\Gamma^{\text{r}}_{\text{int}} ( t )$ and $f_{\text{r}} ( t )$ by measuring each of the two resonators over four separate~$120$-\si{\hour} time periods. The time series data is acquired by cycling over different values of~$P_{\text{p}}$ and $\Delta_{\text{r,p}}$ with a period~$\Delta t = \SI{520}{\second}$. Each resonator measurement is realized by configuring the VNA with an intermediate frequency bandwidth~$\Delta f_\text{IF} = \SI{1}{\hertz}$, $116$ frequency points (including normalization points), and a measurement power of~$- \SI{30}{dBm}$; this power corresponds to~$P_{\text{pr}} \sim - \SI{160}{dBm}$ at the sample. With these settings, each resonator measurement takes~$t_{\text{meas}} \approx \SI{103}{\second}$. The microwave source used to generate the pump field is set to be either \emph{off} or \emph{on} at one of four $2$-tuple of parameters: $( P_{\text{ms}} \, [\si{dBm}] , \Delta_{\text{r,p}} \, [\si{\mega\hertz}] ) \in \{ (-3 , 2),(-3 , -2),(10 , 2),(10 , -2) \}$; we refer to the~\SI{-3}{dBm} settings as \emph{low power} and to the~\SI{10}{dBm} settings as \emph{medium power}. The total attenuation between the source and the sample is~$A \approx \SI{89}{\decibel}$; thus, $P_{\text{p}} = P_{\text{ms}} - A$. Notably, in our experiments a detuning of~\SI{2}{\mega\hertz} is equivalent to approximately~$100$ resonator linewidths; a linewidth is the full width at half maximum of the resonator's Lorentzian curve.

\subsection{Simulations}
	\label{Subsec:METHODS:Simulations}

\begin{table}[b!]
	\caption{Simulation parameters for step~(I). $B_{\text{Q-TLS}}$: Q-TLS bandwidth. $D$: Q-TLS density. $N_{\text{Q-TLS}}$: number of Q-TLSs. $V_{\text{int}}$: interaction region volume. $g_{\text{min}}$: minimum coupling strength cutoff. The parameters~$\tilde{f}_{\text{r}}$, $\Gamma^{\text{r}}_{\text{ext}}$, and $\widetilde{\Gamma}^{\text{r}}_{\text{int}}$ are reported in Table~\ref{Tab:App:Sample:and:characterization:parameters}. All the other parameters required to complete step~(I) are identical to those used in Ref.~\cite{Bejanin:2021} \label{Tab:Simulation:parameters:for:step:(I)}.}
	\begin{center}
	\begin{ruledtabular}
		\begin{tabular}{lr}
			Parameter & Value \\
			\hline
			\raisebox{0mm}[4mm][0mm]{$B_{\text{Q-TLS}}$} (\si{\mega\hertz}) & $300$ \\
			\raisebox{0mm}[3mm][0mm]{$D$} (\si{\per\giga\hertz\per\micro\meter\cubed}) & $400$ \\
			\raisebox{0mm}[3mm][0mm]{$N_{\text{Q-TLS}}$} & $\sim 7500$ \\
			\raisebox{0mm}[3mm][0mm]{$V_{\text{int}}$} (\si{\micro\meter\cubed}) & $157$ \\
			\raisebox{0mm}[3mm][0mm]{$g_{\text{min}}$} (\si{\kilo\hertz}) & $2$ \\
		\end{tabular}
	\end{ruledtabular}
	\end{center}
\end{table}

We compare the time series experiments with simulations similar to those presented in the work of Ref.~\cite{Bejanin:2021}. Here, we expand the scope of that work by simulating not only stochastic time fluctuations in the energy relaxation rate (i.e., $\Gamma^{\text{r}}_{\text{int}}$) but also in~$f_{\text{r}}$. Importantly, we also explore resonator pumping effects with different values of~$( P_{\text{ms}} , \Delta_{\text{r,p}} )$, therefore allowing us to test our GTM-based model under more general conditions.

As in Ref.~\cite{Bejanin:2021}, the procedure to simulate the effect of TLSs on the fluctuations in~$\Gamma^{\text{r}}_{\text{int}}$ and $f_{\text{r}}$ comprises three steps: (I) Generate an ensemble of Q-TLSs interacting with the resonator. (II) Generate several T-TLSs interacting with each Q-TLS. (III) Generate a time series for each T-TLS and propagate the effect of the T-TLSs' switching state to each Q-TLS, and, finally, to the resonator. However, the simulation procedure has to be modified to account for the expanded scope of the present work and for the fact that we are studying resonators instead of a qubit.

In step~(I), the interdigital geometry of the resonator capacitor results in a different electric field~$\vec{E}_{\text{r}}$ compared to the qubit (see App.~\ref{App:RESONATOR:ELECTRIC:FIELD}). The field~$\vec{E}_{\text{r}}$ is needed when determining the coupling strength between the resonator and each Q-TLS, $g = \vec{p} \cdot \vec{E}_{\text{r}} / h$, where~$\vec{p}$ is the Q-TLS electric dipole moment. All the simulation parameters required to complete step~(I) are reported in Table~\ref{Tab:Simulation:parameters:for:step:(I)}.

The parameters used in step~(II) are identical to those reported in Ref.~\cite{Bejanin:2021}.

In step~(III), the simulated time series are generated using Eqs.~(\ref{subeq:Gamma:r:int}) and (\ref{subeq:f:r}), which are based on Eqs.~(\ref{subeq:kappa:rQ-TLS:int}) and (\ref{subeq:2pi:delta:f:r}). In this step, the main departure from Ref.~\cite{Bejanin:2021} is due to the Q-TLS population~$\sigma_{z0}$ of Eq.~(\ref{eq:sigma:z0}). This population depends on the estimated sample temperature~$T \sim \SI{60}{\milli\kelvin}$ and $\langle n \rangle$, where the latter is obtained from~$P_{\text{p}}$ by means of Eq.~(\ref{eq:mean:n})~\footnote{We use~$\tilde{\Gamma}^{\text{r}}_{\text{int}}$ instead of~$\Gamma^{\text{r}}_{\text{int}}$ because, at this stage, we do not have access to~$\Gamma^{\text{r,Q-TLS}}_{\text{int}}$. Given the other quantities in the denominator, this is an acceptable approximation.}. Except for the pump power and detuning, the pump-off and pump-on simulations are identical to each other. In particular, after randomly generating the Q-TLS frequency time series of Eq.~(\ref{eq:fkt}), the same frequency series are used for all different pump~settings.

The collections of T-TLSs and Q-TLSs used in the simulations are randomly drawn from distributions; therefore, each realization of a simulation is unique. In order to match the experimental time series, we run a few simulations and select those that visually resemble the experiments.

\begin{figure*}[ht!]
	\centering
	\includegraphics[width=17.95cm]{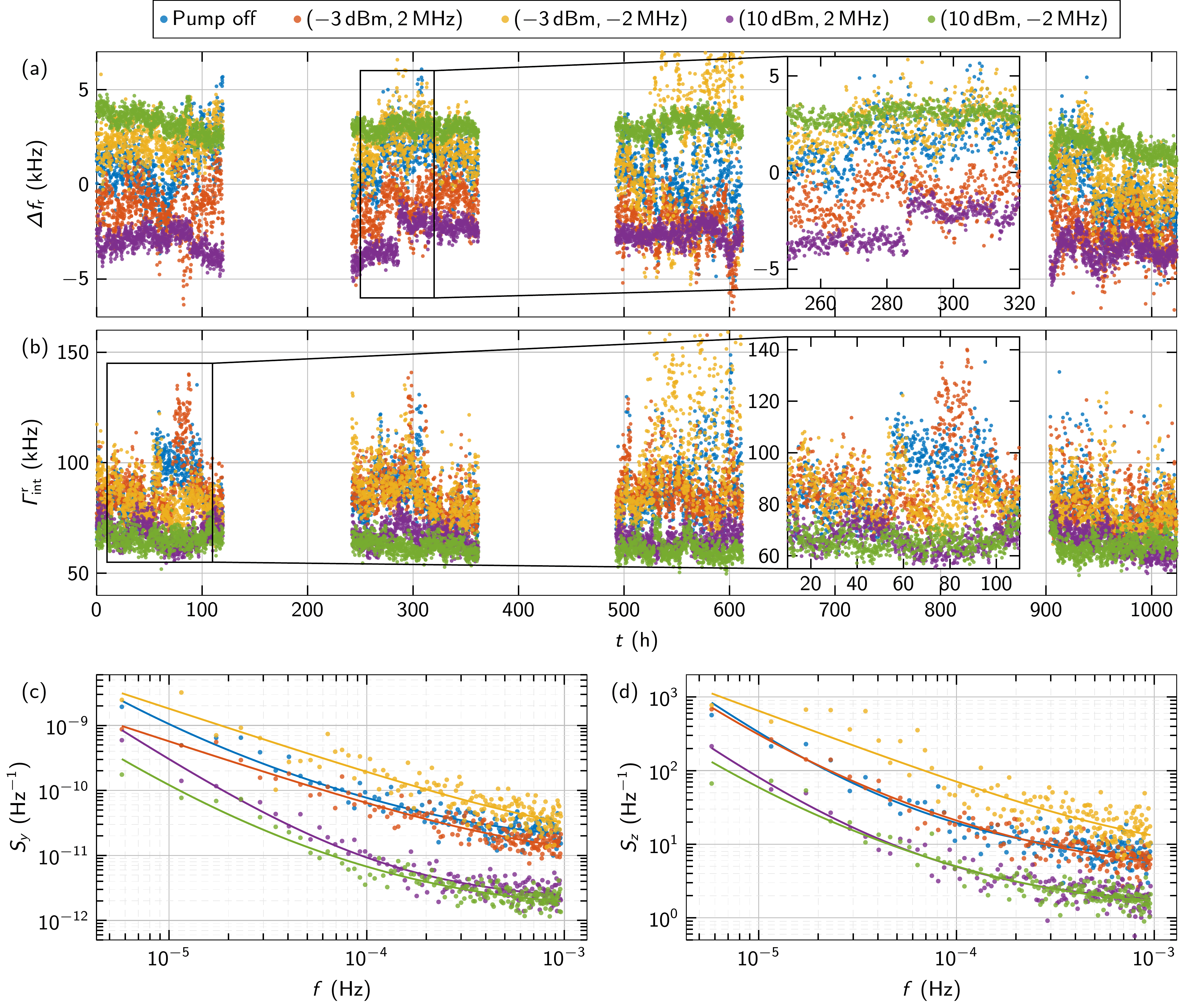}
	\caption{Fluctuation spectroscopy experiments for~R1. (a),(b) $\Delta f_{\text{r}}$ and $\Gamma^{\text{r}}_{\text{int}}$ vs.~$t$; the mean fit errors are~\SI{150}{\hertz} and \SI{1.9}{\kilo\hertz}, respectively. The four~$120$-\si{\hour} time periods are displayed in real time; the blank periods correspond to gaps between experiments. For presentation clarity, we clip the~$y$-axis, therefore loosing a few data points. Insets: detail of selected time windows displaying pronounced telegraphic jumps. (c),(d) $S_y$ and $S_z$ vs.~$f$. Dots correspond to PSD values and solid lines (same color code as dots) to fitting curves from Eq.~(\ref{eq:Sf}). \label{Fig:RESULTS:Fluctuation:spectroscopy:experiments:R1}}
\end{figure*}

\begin{table*}[t!]
	\caption{Fluctuation spectroscopy pump-field settings and parameters extracted from the experimental time series and PSD fits of Figs.~\ref{Fig:RESULTS:Fluctuation:spectroscopy:experiments:R1} (R1) and S1 (R2). We estimate~$\langle n \rangle$ using Eq.~(\ref{eq:mean:n}), where~$\Gamma^{\text{r}}_{\text{ext}} = \widetilde{\Gamma}^{\text{r}}_{\text{ext}}$ from Table~\ref{Tab:App:Sample:and:characterization:parameters} and $\Gamma^{\text{r}}_{\text{tot}} = \widetilde{\Gamma}^{\text{r}}_{\text{ext}} + \langle \Gamma^{\text{r}}_{\text{int}} ( t ) \rangle$; $\langle n \rangle$ for pump off is solely due to~$P_{\text{pr}}$. For the average values~$\langle f_{\text{r}} ( t ) \rangle$ and $\langle \Gamma^{\text{r}}_{\text{int}} ( t ) \rangle$, the number in parenthesis indicates the standard deviation of the associated time series. $h^{y,z}_{0,-1,-2}$: fitting parameters estimated for either~$S_y$ or $S_z$ using Eq.~(\ref{eq:Sf}); standard errors of the fit are shown in parenthesis.
	\label{Tab:RESULTS:Fluctuation:spectroscopy}}
\begin{center}
	\begin{ruledtabular}
		\begin{tabular}{lccccccccccc}
			\raisebox{0mm}[3mm][0mm]{Res.} & $P_{\text{p}}$ & $\Delta_{\text{r,p}}$ & $\langle n \rangle$ & $\langle f_{\text{r}} ( t ) \rangle$ & $\langle \Gamma^{\text{r}}_{\text{int}} ( t ) \rangle$ & $h^y_0$ & $h^y_{-1}$ & $h^y_{-2}$ & $h^z_0$ & $h^z_{-1}$ & $h^z_{-2}$ \\
			{} & (\si{dBm}) & (\si{\mega\hertz}) & (--) & (\si{\giga\hertz}) & (\si{\kilo\hertz}) & (\SI{e-12}{\per\hertz}) & (\num{e-15}) & (\SI{e-18}{\hertz}) & (\si{\per\hertz}) & (\num{e-3}) & (\SI{e-9}{\hertz}) \\
\\[-2.5mm]
\hline\\[-2.5mm]
			\multirow{5}{*}{\shortstack[l]{R1}}
			& off & -- & $\num{7e-02}$ & $5.581779(2)$ & $84(13)$
			& $15(1)$ & $5.7(6)$ & $0.05(2)$ & $5.9(4)$ & $1.1(2)$ & $21(7)$ \\
			& $-3$ & $\phantom{-}2$ & $\num{3e02}$ & $5.581777(1)$ & $85(12)$
			& $8.0(8)$ & $5.6(3)$ & $0$ & $4.3(3)$ & $1.5(2)$ & $15(6)$ \\
			& $-3$ & $-2$ & $\num{3e02}$ & $5.581781(2)$ & $89(24)$
			& $14(3)$ & $18(1)$ & $0$ & $7(1)$ & $6.4(6)$ & $0$ \\
			& $10$ & $\phantom{-}2$ & $\num{6e03}$ & $5.5817760(7)$ & $66(5)$
			& $1.9(2)$ & $0.48(8)$ & $0.03(5)$ & $1.5(1)$ & $0.30(4)$ & $5(2)$ \\
			& $10$ & $-2$ & $\num{6e03}$ & $5.5817818(7)$ & $63(4)$
			& $1.5(1)$ & $0.45(5)$ & $0.007(2)$ & $1.34(8)$ & $0.34(3)$ & $2(1)$ \\
\hline\\[-2.5mm]
			\multirow{5}{*}{\shortstack[l]{R2}}
			& off & -- & $\num{6e-02}$ & $6.081403(2)$ & $118(15)$
			& $19(2)$ & $9.1(9)$ & $0.06(3)$ & $3.6(3)$ & $1.0(1)$ & $27(7)$ \\
			& $-3$ & $\phantom{-}2$ & $\num{3e02}$ & $6.081400(3)$ & $128(21)$
			& $26(3)$ & $7(1)$ & $0.15(5)$ & $5.8(6)$ & $2.5(2)$ & $0$ \\
			& $-3$ & $-2$ & $\num{3e02}$ & $6.081404(2)$ & $132(27)$
			& $22(1)$ & $4.5(6)$ & $0.11(3)$ & $4.8(3)$ & $1.2(2)$ & $38(8)$ \\
			& $10$ & $\phantom{-}2$ & $\num{5e03}$ & $6.0813987(8)$ & $104(7)$
			& $3.4(3)$ & $1.0(1)$ & $0.011(4)$ & $1.14(7)$ & $0.44(3)$ & $0$ \\
			& $10$ & $-2$ & $\num{5e03}$ & $6.0814061(9)$ & $103(7)$
			& $4.0(3)$ & $0.8(1)$ & $0.026(6)$ & $1.18(9)$ & $0.49(3)$ & $0$ \\
[-0.5mm]
		\end{tabular}
	\end{ruledtabular}
\end{center}
\end{table*}

\subsection{Spectral analysis}
	\label{Subsec:METHODS:Spectral:analysis}

We analyze the experimental and simulated time series by estimating and fitting the~PSD. The PSD is estimated using Welch's method
with approximately~$48$-\si{\hour} segments and rectangular windows; we use~\SI{50}{\percent} overlap within each of the four individual~$120$-\si{\hour} time periods and no overlap across separate periods (in total, $4 \times 4 = 16$ segments).

We estimate the PSD for both~$\Gamma^{\text{r}}_{\text{int}} ( t )$ and $f_{\text{r}} ( t )$ by normalizing them as
\begin{displaymath}
y ( t ) = \frac{f_{\text{r}} ( t ) - \langle f_{\text{r}} ( t ) \rangle}{\langle f_{\text{r}} ( t ) \rangle} \qand z ( t ) = \frac{\Gamma^{\text{r}}_{\text{int}} ( t ) - \langle \Gamma^{\text{r}}_{\text{int}} ( t ) \rangle}{\langle \Gamma^{\text{r}}_{\text{int}} ( t ) \rangle} .
\end{displaymath}

The PSD noise model, which can be used to analyze either~$y ( t )$ or $z ( t )$, reads as
\begin{equation}
S ( f ) = h_0 + \frac{h_{-1}}{f} + \frac{h_{-2}}{f^2} ,
	\label{eq:Sf}
\end{equation}
where~$f$ is the analysis frequency and $h_0$, $h_{-1}$, and $h_{-2}$ are the amplitudes of the white noise, $1/f$ (flicker) noise, and random walk (Brownian) noise, respectively.

We choose to fit~$\log_{10} [ S ( f ) ]$ using the Levenberg-Marquardt algorithm; the fitting parameters are~$h_0$, $h_{-1}$, and $h_{-2}$. The logarithm eliminates the magnitude difference between high- and low-frequency noise, allowing for an evenly weighted fit~\footnote{More precisely, we fit~$\log_{10} [ S ( f ) / s_0 ]$, where the scaling factor~$s_0 = 10^{\round{\log_{10} \tilde{h}_0}}$; $\tilde{h}_0$ is the initial guess for~$h_0$ and $\round{\cdot}$ is the rounding function.}. Additionally, the fitting parameters are lower bounded to zero.

\section{RESULTS}
	\label{Sec:RESULTS}

The main results of this work are presented in Fig.~\ref{Fig:RESULTS:Fluctuation:spectroscopy:experiments:R1}, which
shows the experimental time series~$\Gamma^{\text{r}}_{\text{int}} ( t )$ and $f_{\text{r}} ( t )$ as well as the PSDs of the normalized time series~$y ( t )$ and $z ( t )$, $S_y ( f )$ and $S_z ( f )$, for~R1. Similar results for~R2 are shown in Fig.~S1 of the Supplemental Material. A set of parameters extracted from the experimental time series and PSD fits for both~R1 and R2 are reported in Table~\ref{Tab:RESULTS:Fluctuation:spectroscopy}. We note that, at low and medium power, $\langle n \rangle$ is approximately~$4000$ and $90000$ times larger than when the pump is off.

A continuous~$120$-\si{\hour} time period would allow us to reach~$f = \SI{2.3e-6}{\hertz}$; however, we choose to trade frequency bandwidth at low frequency for accuracy by averaging our data with Welch's method. We average over~$16$ time windows, resulting in a fourfold reduction in the PSD variance. This approach makes it possible to accurately characterize fluctuations at very low frequencies.

A visual inspection of Fig.~\ref{Fig:RESULTS:Fluctuation:spectroscopy:experiments:R1}~(a) reveals that~$f_{\text{r}}$ is pulled towards~$f_{\text{p}}$, as in Fig.~\ref{Fig:METHODS:Pumped:resonator:characterization}~(a). A similar inspection of Fig.~\ref{Fig:RESULTS:Fluctuation:spectroscopy:experiments:R1}~(b) indicates that~$\Gamma^{\text{r}}_{\text{int}}$ remains practically unchanged when the pump field is either off or at low power, whereas it is reduced at medium power, as in Fig.~\ref{Fig:METHODS:Pumped:resonator:characterization}~(d). These qualitative findings are corroborated by the average values~$\langle f_{\text{r}} ( t ) \rangle$ and $\langle \Gamma^{\text{r}}_{\text{int}} ( t ) \rangle$ in Table~\ref{Tab:RESULTS:Fluctuation:spectroscopy}. Importantly, the noise level of the medium-power traces is noticeably reduced compared to the off and low-power regimes (in Fig.~S3, we show that, at high power, the noise is even further reduced).

Performing measurements at different pump-field settings unveils a striking characteristic in the time series: Traces associated with different values of~$P_{\text{p}}$ and $\Delta_{\text{r,p}}$ do not track each other. We refer to this behavior as the \emph{time-series asymmetry}. For example, the inset of Fig.~\ref{Fig:RESULTS:Fluctuation:spectroscopy:experiments:R1}~(a) shows that~$\Delta f_{\text{r}}$ undergoes a pronounced telegraphic jump at~$t \approx \SI{285}{\hour}$ for~$(\SI{10}{dBm} , \SI{2}{\mega\hertz})$ (purple trace) but not for any other settings. The inset of Fig.~\ref{Fig:RESULTS:Fluctuation:spectroscopy:experiments:R1}~(b) evinces a similar behavior at~$t \approx \SI{50}{\hour}$ for pump off (blue trace) and at $t \approx \SI{70}{\hour}$ for~$(\SI{-3}{dBm} , \SI{2}{\mega\hertz})$ (orange trace). One more example is the time series for~$(\SI{-3}{dBm} , \SI{-2}{\mega\hertz})$ (yellow trace), which shows drastic amplitude fluctuations in the third time period for both~$f_{\text{r}}$ and $\Gamma^{\text{r}}_{\text{int}}$. We emphasize that our data is collected by cycling the pump-field settings (see Sec.~\ref{Sec:METHODS}) and, thus, such large variations among different traces are unexpected.

The PSDs shown in Figs.~\ref{Fig:RESULTS:Fluctuation:spectroscopy:experiments:R1}~(c) and (d) confirm the behavior observed in the time series. Notably, we measure~\mbox{$\sim 1/f$-noise} persisting down to~$f = \SI{5.8e-6}{\hertz}$. At medium power, as expected, the overall noise level is significantly reduced compared to pump off; unexpectedly, the noise level at low power is similar or even slightly higher. We refer to this overall behavior as \emph{noise-level power scaling}. The yellow trace [$(\SI{-3}{dBm} , \SI{-2}{\mega\hertz})$], for example, displays drastic time-series fluctuations that lead to a higher noise level in~$S_y$ and $S_z$. We also notice another effect: The yellow trace exhibits a higher noise level than the orange trace [$(\SI{-3}{dBm} , \SI{2}{\mega\hertz})$], despite~$P_{\text{p}}$ being the same for both traces. Additionally, we notice that the~$S_y$ noise level for the purple trace [$(\SI{10}{dBm} , \SI{2}{\mega\hertz})$] is slightly higher than for the green trace [$(\SI{10}{dBm} , \SI{-2}{\mega\hertz})$]. We refer to this effect as \emph{noise-level asymmetry}.

\begin{figure}[t!]
	\centering
	\includegraphics[width=\columnwidth]{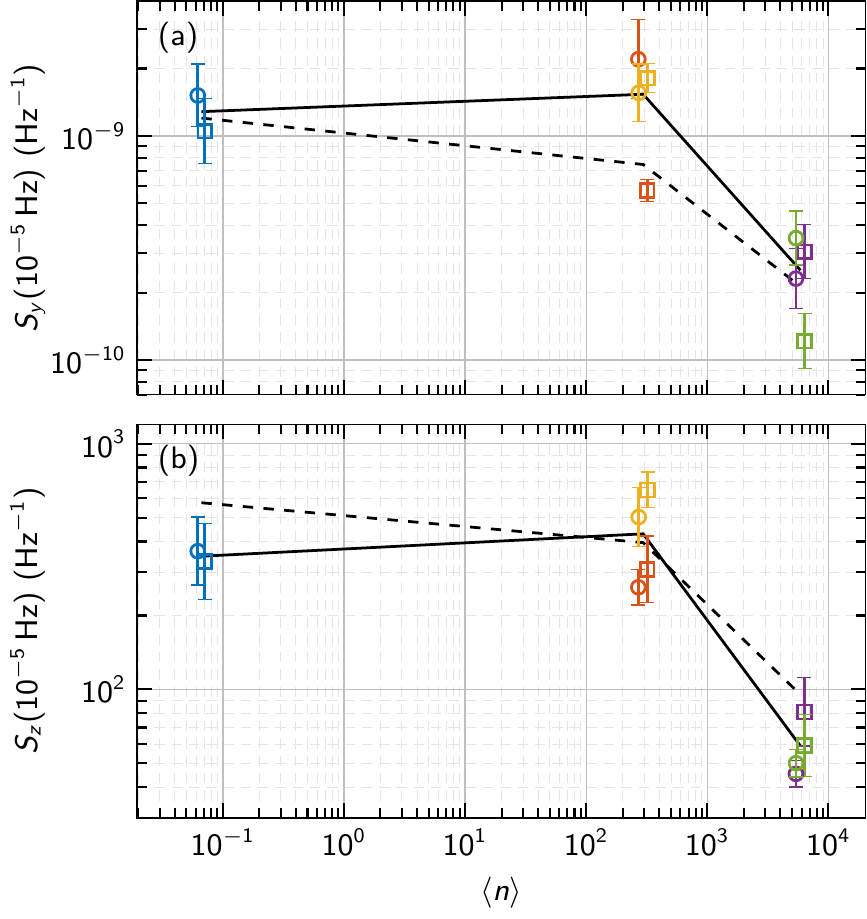}
	\caption{Fluctuation spectroscopy summary. (a),(b) $S_y$ and $S_z$ at~$f = \SI{e-5}{\hertz}$ vs.~$\langle n \rangle$ for~R1 (squares) and R2 (circles); the color code is the same as in Fig.~\ref{Fig:RESULTS:Fluctuation:spectroscopy:experiments:R1}. The PSD values are from the fitting curves in Fig.~\ref{Fig:RESULTS:Fluctuation:spectroscopy:experiments:R1}. Error bars represent~\SI{95}{\percent} prediction intervals of the curve value. The solid lines indicate the overall power-scaling trend. The dashed lines are obtained from averaging the simulated noise levels for~R1 and R2.
	\label{Fig:RESULTS:Fluctuation:spectroscopy:summary}}
\end{figure}

It is worth stressing that all the features observed for~R1, particularly the time-series asymmetry, are also observed in the time series and PSDs for~R2 (see Fig.~S1), suggesting these effects are reproducible. We summarize the fluctuation spectroscopy results for~R1 and R2 in Fig.~\ref{Fig:RESULTS:Fluctuation:spectroscopy:summary}, which displays~$S_y$ and $S_z$ at very low frequency for the five pump-field settings investigated in this work. This figure highlights our findings on noise-level power scaling and asymmetry.

In our work of Ref.~\cite{Bejanin:2021}, we have already demonstrated that GTM-based simulations can match~$\Gamma^{\text{r}}_{\text{int}} ( t )$ for an Xmon transmon qubit; however, in that study we were unable to investigate fluctuations in~$f_{\text{r}} ( t )$. In fact, Xmon qubits are highly susceptible to flux noise, making them effectively insensitive to TLS fluctuations in the frequency. Measuring and simulating resonators, as in the present work, allows us to analyze TLS-induced fluctuations also in~$f_{\text{r}}$ and, thus, to further corroborate the GTM.

\begin{figure*}[ht!]
	\centering
	\includegraphics[width=17.95cm]{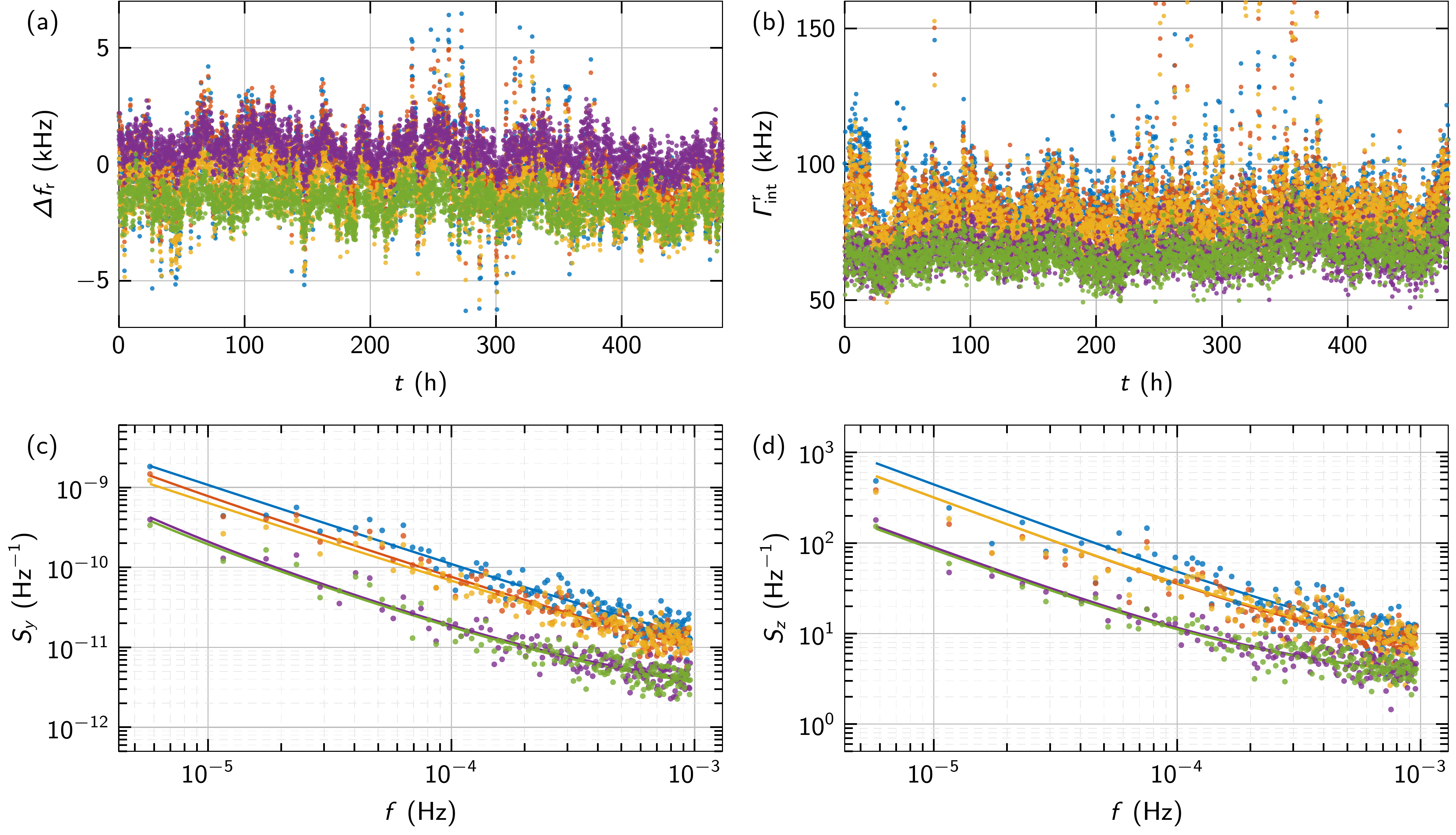}
	\caption{Fluctuation spectroscopy simulations for~R1. (a),(b) $\Delta f_{\text{r}}$ and $\Gamma^{\text{r}}_{\text{int}}$ vs.~$t$; the simulations are performed for~\SI{480}{\hour} and split into four~$120$-\si{\hour} time periods for the spectral analysis. For presentation clarity, we clip the~$y$-axis, therefore loosing a few data points. (c),(d) $S_y$ and $S_z$ vs.~$f$. Dots correspond to PSD values and solid lines to fitting curves from Eq.~(\ref{eq:Sf}). The color code is the same as in Fig.~\ref{Fig:RESULTS:Fluctuation:spectroscopy:experiments:R1}. \label{Fig:RESULTS:Fluctuation:spectroscopy:simulations:R1}}
\end{figure*}

Figure~\ref{Fig:RESULTS:Fluctuation:spectroscopy:simulations:R1} shows simulations (see Sec.~\ref{Subsec:METHODS:Simulations}) of the experiments reported in Fig.~\ref{Fig:RESULTS:Fluctuation:spectroscopy:experiments:R1}; similar simulations for~R2 are shown in Fig.~S2. When the pump is off, the simulations accurately reproduce the experiments for both~$f_{\text{r}}$ and $\Gamma^{\text{r}}_{\text{int}}$. When the pump is set to medium power, the simulations also capture the overall reduction in~$\Gamma^{\text{r}}_{\text{int}}$ as well as a reduction in the noise level for both~$S_y$ and $S_z$. In addition, the simulated~$f_{\text{r}}$ is generally pulled towards~$f_{\text{p}}$. Figure~\ref{Fig:RESULTS:Fluctuation:spectroscopy:summary} shows that, at low frequency (\SI{e-5}{\hertz}), the simulated average noise levels match fairly well the experiments at all powers.

Interestingly, we observe differences between simulations and experiments at medium and, more prominently, at low power. Firstly, the simulated~$f_{\text{r}} ( t )$ and $\Gamma^{\text{r}}_{\text{int}} ( t )$ time series for pump on track the pump-off series. The simulated time series resemble scaled versions of each other; in contrast, the experimental traces are characterized by a time-series asymmetry. Secondly, at low power, the simulated noise level is reduced compared to pump off. This behavior is clearly expected in the simulations because any amount of Q-TLS saturation necessarily leads to a monotonic reduction in the impact of Q-TLSs [see Eqs.~(\ref{subeq:Gamma:r:int}) and (\ref{subeq:f:r})]. Although experiments and simulations agree fairly well in this regard (Fig.~\ref{Fig:RESULTS:Fluctuation:spectroscopy:summary}), it is for this reason that the slight increase in experimental noise level at low power is unexpected. Finally, the simulations do not show any noise-level asymmetry; instead, they reveal an almost identical noise level for different values of~$\Delta_{\text{r,p}}$, at any given~$P_{\text{p}}$.

\section{DISCUSSION}
	\label{Sec:DISCUSSION}

It is comforting that the theoretical model described in Sec.~\ref{Sec:THEORY} allows us to explain the experimental results when the pump is off as well as the general noise trend at low frequency, as explained in Sec.~\ref{Sec:RESULTS}. At low and medium power, however, the departures between experiments and simulations suggest that injecting photonic excitations into the system results in more complex dynamics. Such dynamics are not entirely captured by the independent Jaynes-Cummings model implemented in our simulations.

The low and medium power regimes correspond to a scenario where a few hundred to a few thousand photons populate a system with approximately~$10000$ lossy Q-TLSs, which are \emph{simultaneously} coupled to one resonator. A more sophisticated model is likely needed to capture the complexity of this system, such as a driven-dissipative Tavis-Cummings model for the Q-TLS--resonator interaction, where, following the GTM, each Q-TLS also interacts with a few T-TLSs undergoing stochastic fluctuations. At high power, the Q-TLSs are effectively turned off by saturation, thereby leading to simpler dynamics. It may also be required to include the effect of Q-TLS--Q-TLS interactions~\cite{Burnett:2014}. Unfortunately, a quantum computer would be likely required to simulate a Hilbert space of this size.

In the works of Refs.~\cite{Burnett:2014,DeGraaf:2018,Niepce:2021}, the authors have measured~$f_{\text{r}} ( t )$ up to approximately~\SI{3}{\hour}. While their measurement method allows them to access higher frequencies than ours, their lowest frequency is limited to larger than~$\SI{e-3}{\hertz}$. Additionally, their measurements are performed for~$\langle n \rangle \gtrsim 2$ in Refs.~\cite{Burnett:2014,DeGraaf:2018} and $\gtrsim 40$ in Ref.~\cite{Niepce:2021} (i.e., higher than ours) and only at~$\Delta_{\text{r,p}} = 0$ (i.e., only on resonance). Within these constraints, their PSDs generally indicate a progressive noise reduction with increasing~$\langle n \rangle$. These results are consistent with our observations at medium and high power (see Fig.~S3 for the high power measurements) but cannot be compared to our findings at low power and pump off. Interestingly, the time series reported in Ref.~\cite{Niepce:2021} display a single dominant TLS in all time series; in contrast, our data reveals the presence of an ensemble of Q-TLSs. It is also worth noting that their technique does not make it possible to measure~$\Gamma^{\text{r}}_{\text{int}}$, which, in our experiments, allows us to gain further information about the resonators' noise processes.

\section{CONCLUSIONS}
	\label{Sec:CONCLUSIONS}

In this work, we measure long time series of~$f_{\text{r}}$ and $\Gamma^{\text{r}}_{\text{int}}$, allowing us to accurately characterize noise at very low frequency. In summary, our main results are:
\begin{enumerate}[(1)]
\item Time-series asymmetry; that is, time series do not track each other.
\item Noise-level asymmetry; that is, for a given~$P_{\text{p}}$, $\Delta_{\text{r,p}}$ affects the PSDs' noise level.
\item Noise-level power scaling; that is, the PSDs' noise level is unchanged at low power and significantly reduced at medium power. Additionally, any Q-TLS--induced noise is almost entirely suppressed at high power.
\end{enumerate}

When comparing our experiments to simulations, we find a reasonable agreement but also a few interesting discrepancies. These indicate that further investigations are required to explain all the features observed in the experiments. Future work may focus on added complexity to the model, as suggested in Sec.~\ref{Sec:DISCUSSION}, or an entirely new approach to the description of stochastic time fluctuations due to driven-dissipative TLS interactions.

Finally, an outlook on possible quantum computing applications. Pumping resonators off resonance leads to improvements in~$Q^{\text{r}}_{\text{int}}$~\cite{Sage:2011,Kirsh:2017,Capelle:2020}. In Fig.~S3, we show the results of a high-power pump at~$\Delta_{\text{r,p}} = \SI{-2}{\mega\hertz}$. When the pump is on, $S_y(\SI{e-5}{\hertz})$ and $S_z(\SI{e-5}{\hertz})$ are reduced by a factor of~$\sim 194$ and $133$, respectively (we find similar results on other three resonators; data not shown). Naively, one may think to apply such a pumping to qubits in order to increase their~$T_1$ and reduce time fluctuations; however, a strong pump, even when detuned by~$\sim 100$ linewidths from the qubit transition frequency, results in a significant qubit population. Recently, an alternative method based on low-frequency Landau-Zener transitions has been shown to improve~$Q^{\text{r}}_{\text{int}}$, while keeping the resonator in the vacuum state~\cite{Matityahu:2019}. Our results of Fig.~S3 indicate that TLS saturation can be used as a powerful tool to largely improve qubit operations. Using \emph{low-frequency} Landau-Zener transitions to reach saturation could, thus, lead to a major breakthrough in the reduction not only of loss but also of noise in superconducting qubits.

\begin{acknowledgments}
This research was undertaken thanks in part to funding from the Canada First Research Excellence Fund~(CFREF). We acknowledge the support of the Natural Sciences and Engineering Research Council of Canada~(NSERC), [Application Number: RGPIN-2019-04022]. We would like to acknowledge the Canadian Microelectronics Corporation~(CMC) Microsystems for the provision of products and services that facilitated this research, including CAD and ANSYS, Inc software. The authors thank the Quantum-Nano Fabrication and Characterization Facility at the University of Waterloo, where the sample was fabricated.
\end{acknowledgments}

\appendix

\section{Q-TLS PARTIAL CONTRIBUTIONS}
	\label{App:Q-TLS:PARTIAL:CONTRIBUTIONS}

Following the derivation in the work of Ref.~\cite{Capelle:2020}, Eq.~(\ref{eq:rhodot}) allows us to find the quantized Maxwell-Bloch equations for the resonator field~$\langle \hat{a} \rangle$, Q-TLS coherence~$\langle \hat{\sigma} \rangle$, and Q-TLS population~$\langle \hat{\sigma}_z \rangle$. In the \emph{stationary regime}, these equations make it possible to find~\mbox{$\langle \hat{\sigma}_z \rangle = \sigma_{z0}$}:
\begin{equation}
\sigma_{z0} = \dfrac{\left( \Gamma^{\text{Q-TLS}}_1 \right)^{\!\!2} \dfrac{\langle n \rangle}{n_{\text{s}}}}{16 \pi^2 \Delta^2_{\text{Q-TLS,p}} \! + \! \left( \Gamma^{\text{Q-TLS}}_1 \right)^{\!\!2} \! \left( \! 1 + \dfrac{\langle n \rangle}{n_{\text{s}}} \! \right)} - 1 ,
	\label{eq:sigma:z0}
\end{equation}
where~$1 / n_{\text{s}} = 32 \pi^2 g^2 / \left( \Gamma^{\text{Q-TLS}}_1 \right)^{\!\!2}$; $\langle n \rangle$ is generated by steadily pumping the resonator (see Sec.~\ref{Subsec:METHODS:Off-resonant:pumping} for details), and Q-TLS saturation is reached upon exceeding~$n_{\text{s}}$. Equation~(\ref{eq:sigma:z0}) is the saturation law for the population of a single Q-TLS for any~$\Delta_{\text{Q-TLS,p}}$.

In order to find an expression for~$\kappa^{\text{r,Q-TLS}}_{\text{int}}$, we need to consider the \emph{transient regime} of~$\langle \hat{a} \rangle$. This regime is described by a decaying sinusoidal function~\mbox{$\delta \alpha ( t ) = \exp( -\Gamma^{\text{r}}_{\text{ext}} t / 2 ) \exp( i \, \delta \omega \, t )$}, where~$\delta \omega = i \kappa^{\text{r,Q-TLS}}_{\text{int}} / 2 + 2 \pi \, \delta \! f_{\text{r}}$. In the weak coupling approximation, $g \ll \Gamma^{\text{Q-TLS}}_1$, the Maxwell-Bloch equations for the transient dynamics lead to
\begin{subequations}
	\begin{empheq}[]{align}
		\kappa^{\text{r,Q-TLS}}_{\text{int}} & = -\sigma_{z0} \dfrac{16 \pi^2 g^2 \, \Gamma^{\text{Q-TLS}}_1}{16 \pi^2 \Delta^2_{\text{r,Q-TLS}} \! + \! \left( \Gamma^{\text{Q-TLS}}_1 \right)^{\!\!2}}
	\label{subeq:kappa:rQ-TLS:int}
	\end{empheq}
\end{subequations}
and
\setcounter{equation}{\value{equation}-1}
\begin{subequations}
\setcounter{equation}{1}
	\begin{empheq}[]{align}
		 2 \pi \, \delta \! f_{\text{r}} & = \sigma_{z0} \dfrac{16 \pi^2 g^2 \Delta_{\text{r,Q-TLS}}}{16 \pi^2 \Delta^2_{\text{r,Q-TLS}} \! + \! \left( \Gamma^{\text{Q-TLS}}_1 \right)^{\!\!2}} ,
	\label{subeq:2pi:delta:f:r}
	\end{empheq}
\end{subequations}
where~$\Delta_{\text{r,Q-TLS}} = \tilde{f}_{\text{r}} - f_{\text{Q-TLS}} ( t )$.

\section{SAMPLE AND SETUP}
	\label{App:SAMPLE:AND:SETUP}

\begin{figure}[t!]
	\centering
	\includegraphics[width=\columnwidth]{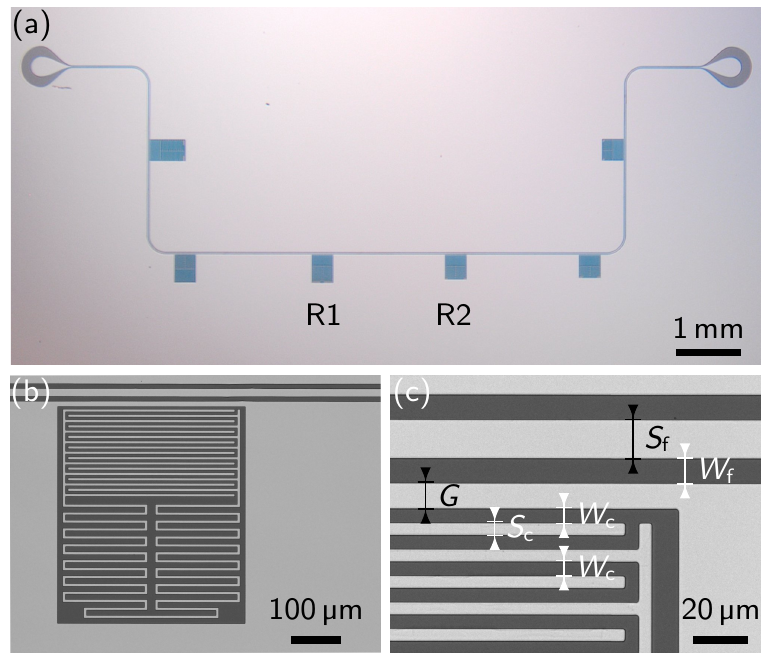}
		\caption{Sample micrographs. (a) Six quasilumped element resonators coupled to a feed line. The measured resonators are labeled~R1 and R2. The sample's input (left) and output (right) pads mate with the three-dimensional wires of a quantum socket~\cite{Bejanin:2016} and allow us to measure~$S_{21}$. (b) Closeup of~R1, showing the capacitive coupling to the feed line (top region) as well as the dc floating ground. (c) Closeup of~R1's top region; the characteristic dimensions (symbols) are reported in the appendix text. Scale markers are indicated in all panels.
	\label{Fig:App:Sample:micrographs}}
\end{figure}

Figure~\ref{Fig:App:Sample:micrographs} shows micrographs of a sample identical to the one measured in this work. The sample is made from a~$100$-\si{\nano\meter}-thick Al film deposited by means of electron-beam evaporation on a Si substrate. The fabrication and substrate cleaning processes are similar to those outlined in our work of Ref.~\cite{Earnest:2018} but without the thermal annealing step.

The quasilumped elements used to implement the resonators are a meandering strip inductor and an interdigital capacitor. The capacitor has a total of~$17$ fingers, where each finger is~$327.5$-\si{\micro\meter} long and has a width~$S_{\text{c}} = \SI{5}{\micro\meter}$ and a gap~$W_{\text{c}} = \SI{5}{\micro\meter}$. The resonance frequency~$\tilde{f}_{\text{r}}$ is set by the inductor's strip length, which varies between~$4431$ and \SI{5375}{\micro\meter}; the strip width of the inductor is~\SI{5}{\micro\meter}, with a minimum intermeander spacing of~\SI{10}{\micro\meter} to avoid parasitic capacitances. The width and gap of the feed line are~$S_{\text{f}} = \SI{15}{\micro\meter}$ and $W_{\text{f}} = \SI{9}{\micro\meter}$, respectively, resulting in~$Z_0 \approx \SI{50}{\ohm}$; the width of the ground plane separating the feed line from the top edge of each resonator is~$G = \SI{10}{\micro\meter}$.

\begin{table*}[t!]
	\caption{Sample and characterization parameters for~R1 and R2. $\ell_{\text{L}}$: inductor's strip length. $L_{\text{r}}$, $C_{\text{r}}$, and $C_{\text{ext}}$ are simulated using Ansys Maxwell by ANSYS, Inc. $\widetilde{\Gamma}^{\text{r}}_{\text{ext}} = 2 \pi \tilde{f}_{\text{r}} / \widetilde{Q}^{\text{r}}_{\text{ext}}$: High-power energy relaxation rate; $\widetilde{Q}^{\text{r}}_{\text{ext}}$ is the high-power external quality factor. Note that~$\widetilde{\Gamma}^{\text{r}}_{\text{ext}}$ is expected to remain approximately constant with~$P_{\text{pr}}$. Using the simulated value of~$C_{\text{ext}}$, it is also possible to estimate~$\widetilde{\Gamma}^{\text{r}}_{\text{ext}} = C^2_{\text{ext}} \tilde{f}^2_{\text{r}} Z_0 / C_{\text{r}}$ within~$\sim \SI{15}{\percent}$ from the measured values. Standard errors of the fit are shown in parenthesis. \label{Tab:App:Sample:and:characterization:parameters}}
	\begin{center}
	\begin{ruledtabular}
		\begin{tabular}{ccccccccccc}
			\raisebox{0mm}[4mm][0mm]{Resonator} & $\tilde{f}_{\text{r}}$ & $\ell_L$ & $L_{\text{r}}$ & $C_{\text{r}}$ & $C_{\text{ext}}$ & $\widetilde{\Gamma}^{\text{r}}_{\text{ext}}$ & $\Gamma^0_{\text{Q-TLS}}$ & $\widetilde{\Gamma}^{\text{r}}_{\text{int}}$ & $n_{\text{c}}$ & $\alpha$ \\
			\raisebox{0mm}[0mm][2mm]{} & {\footnotesize{(\si{\giga\hertz})}} & \footnotesize{(\si{\micro\meter})} & \footnotesize{(\si{\nano\henry})} & \footnotesize{(\si{\femto\farad})} & \footnotesize{(\si{\femto\farad})} & \footnotesize{(\si{\kilo\hertz})} & \footnotesize{(\si{\kilo\hertz})} & \footnotesize{(\si{\kilo\hertz})} & \footnotesize{--} & \footnotesize{--} \\
			\hline
			\hline
			\\[-3.0mm]
			\raisebox{0mm}[3mm][0mm]{R1} & $5.581779$ & $6225$ & $2.68$ & $286$ & $10$ & $584$ & $88.2(2)$ & $21.0(5)$ & $0.07(1)$ & $0.234(9)$ \\
			\raisebox{0mm}[3mm][0mm]{R2} & $6.081402$ & $5214$ & $2.29$ & $286$ & $10$ & $547$ & $113.3(3)$ & $41.6(9)$ & $0.039(9)$ & $0.21(1)$ \\
		\end{tabular}
	\end{ruledtabular}
	\end{center}
\end{table*}

\begin{figure}[b!]
	\centering
	\includegraphics{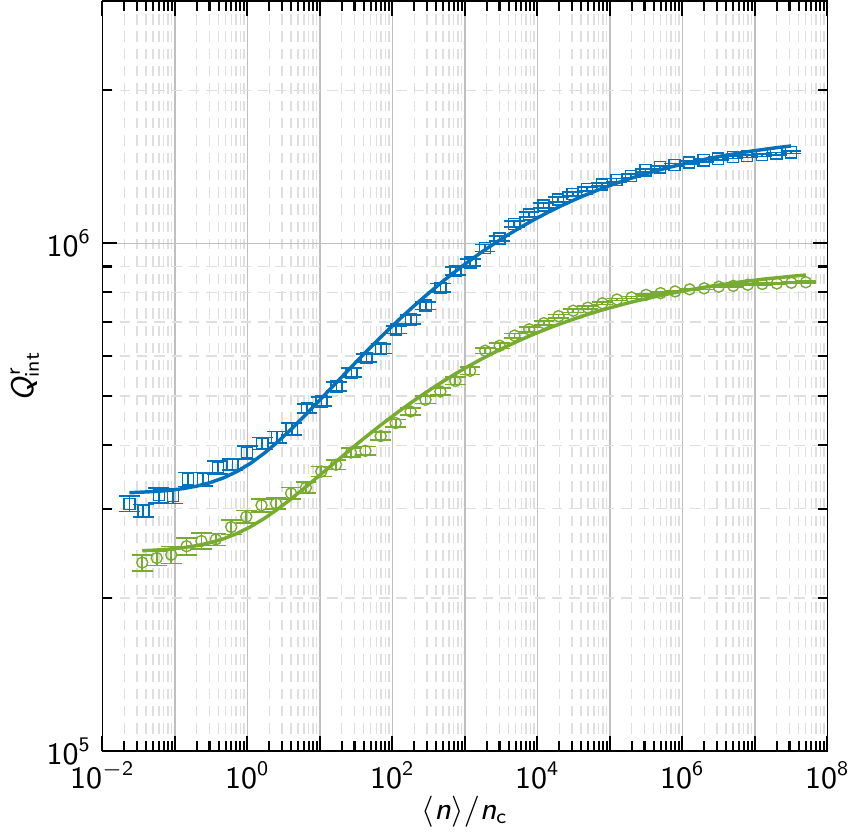}
	\caption{S-curve measurements for~R1 (blue squares) and R2 (green circles). $Q^{\text{r}}_{\text{int}}$ vs.~$\langle n \rangle / n_{\text{c}}$. Solid lines correspond to fitting curves obtained from Eq.~(\ref{eq:inv:Qrint}). All measurements are performed at~$T \sim \SI{10}{\milli\kelvin}$. The fitting parameters are reported in Table~\ref{Tab:App:Sample:and:characterization:parameters}. Error bars represent~\SI{95}{\percent} confidence intervals. \label{Fig:App:S-curve:measurements}}
\end{figure}

Figure~\ref{Fig:App:S-curve:measurements} shows the S-curve measurements for~R1 and R2. When~$T \rightarrow 0^+$, these curves follow the modified STM model~\cite{Earnest:2018}
\begin{equation}
\frac{1}{Q^{\text{r}}_{\text{int}}} = F \tan \delta^0_{\text{Q-TLS}} \left( 1 + \frac{\langle n \rangle}{n_{\text{c}}} \right)^{\!\! - \alpha} + \frac{1}{\widetilde{Q}^{\text{r}}_{\text{int}}} ,
	\label{eq:inv:Qrint}
\end{equation}
where~$F \tan \delta^0_{\text{Q-TLS}} = \Gamma^0_{\text{Q-TLS}} / 2 \pi \tilde{f_{\text{r}}}$ ($F$ is the filling factor for the Q-TLS regions), $\alpha$ is an exponent indicating the deviation from the STM (in the STM, $\alpha = 1 / 2$), and the constant offset~$1 / \widetilde{Q}^{\text{r}}_{\text{int}} = \widetilde{\Gamma}^{\text{r}}_{\text{int}} / 2 \pi \tilde{f}_{\text{r}}$ accounts for all non-TLS losses.

In our experimental setup, the probe field is generated by means of a VNA from Keysight Technologies Inc., model PNA-X N5242A. The pump field is generated by means of a Keysight microwave source, model E8257D-UNY PSG with enhanced ultra-low phase noise. For all measurements, we use a rubidium timebase from Stanford Research Systems, model DG645 opt.~5, to ensure the long-term frequency stability of both probe and pump fields.

The probe and pump fields are superposed with a two-way power combiner from Krytar, Inc., model 6005180-471. As shown in Fig.~S4, the combined field is heavily attenuated and filtered before reaching the sample. The total attenuation of the input line, including power combiner and cables' attenuation, is~$A \approx \SI{89}{\decibel}$, while the gain of the output line is approximately~\SI{59}{\decibel}. The sample is housed in a quantum socket~\cite{Bejanin:2016} anchored to the mixing chamber stage of a dilution refrigerator, at approximately~\SI{10}{\milli\kelvin}. A schematic diagram of the entire experimental setup is shown in Fig.~S4.

\section{RESONATOR ELECTRIC FIELD}
	\label{App:RESONATOR:ELECTRIC:FIELD}

\begin{figure}[bh!]
	\centering
	\includegraphics{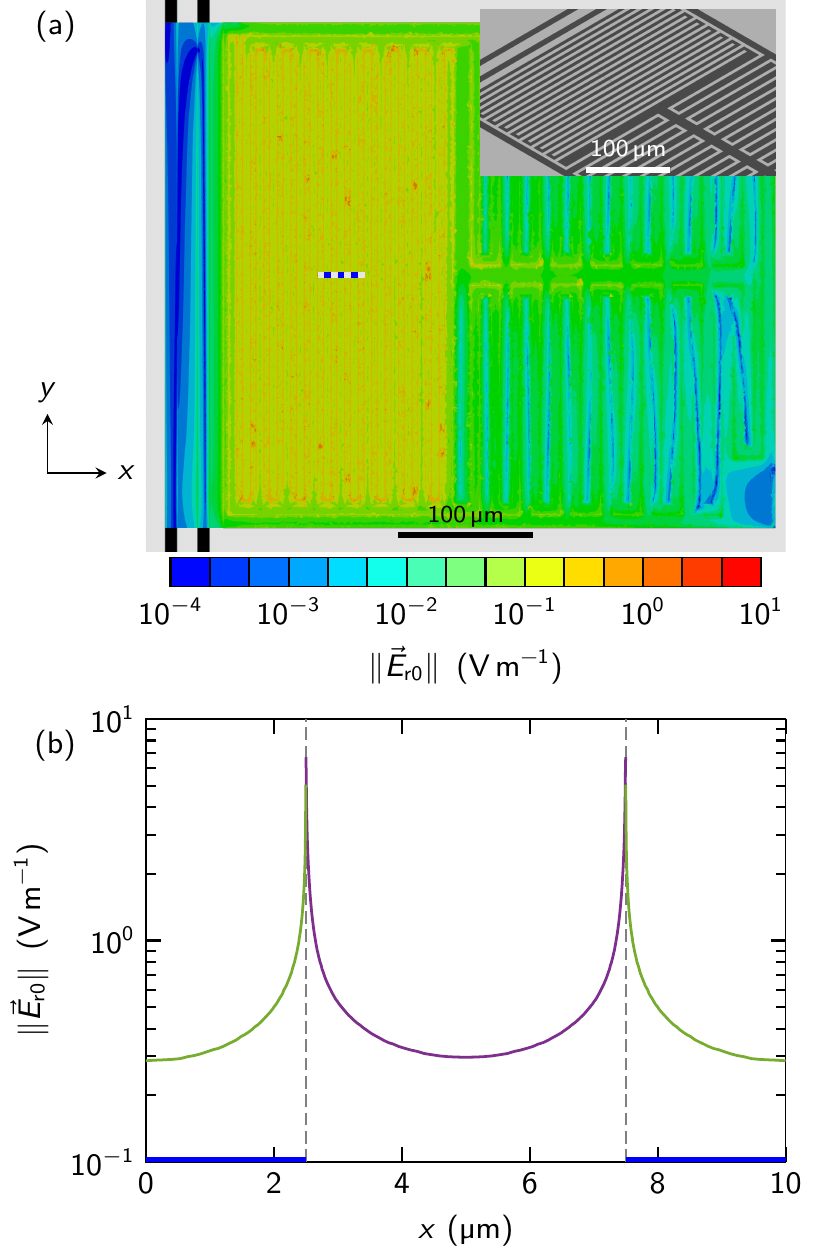}
	\caption{Electric field simulations for~R2. (a) Heat map of~$\norm*{\vec{E}_{\text{r}0}}$ over the entire resonator structure. Inset: three-dimensional view of the resonator capacitor center region. The dashed horizontal blue-white line corresponds to the~$xz$ cross section where~$\vec{E}_{\text{r}}$ is calculated with higher resolution. Blue dashes: regions above the Al films. White dashes: regions above the Si substrate. (b) $\norm*{\vec{E}_{\text{r}0}}$ vs.~$x$. The field is plotted from the midpoint of one of the three fingers within the~$xz$ cross section to the midpoint of the next finger. The thick blue lines indicate Al films. The green portion of~$\norm*{\vec{E}_{\text{r}0}}$ is at~$z = \SI{1.5}{\nano\meter}$ above the~$100$-\si{\nano\meter}--thick Al films, whereas the purple portion is at~$z = \SI{1.5}{\nano\meter}$ above the Si substrate. The electric field variation in proximity of the dashed black vertical lines is approximately~\SI{40}{\percent} between~$z = 0.5$ and \SI{3.0}{\nano\meter}; away from any Al edge, the variation with~$z$ is negligible. Scale markers are indicated in all panels. \label{Fig:App:Electric:field:simualtions}}
\end{figure}

We simulate the electric field~$\vec{E}_{\text{r}}$ for~R2 (same results for~R1) by means of Ansys HFSS by ANSYS, Inc; the geometry details are given in App.~\ref{App:SAMPLE:AND:SETUP} and the layout is shown in Figs.~\ref{Fig:App:Sample:micrographs}~(b) and (c). The simulated electric field is plotted in Fig.~\ref{Fig:App:Electric:field:simualtions}.

Using HFSS, we compute the electric mode volume associated with~R2 (or R1) from
\begin{equation}
V_{\text{m}} = \frac{\varepsilon_{\text{Si}} \iiint_{\Omega_{\text{Si}}} dV \vec{E}_{\text{r}} \cdot \vec{E}^{\ast}_{\text{r}} + \varepsilon_0 \iiint_{\Omega_{\text{v}}} dV \vec{E}_{\text{r}} \cdot \vec{E}^{\ast}_{\text{r}}}{\max \left( \varepsilon_{\text{Si}} \vec{E}_{\text{r,Si}} \cdot \vec{E}^{\ast}_{\text{r,Si}} \, , \, \varepsilon_0 \vec{E}_{\text{r,v}} \cdot \vec{E}^{\ast}_{\text{r,v}} \right)} ,
	\label{eq:Vm}
\end{equation}
where~$\varepsilon_{\text{Si}}$ and $\varepsilon_0$ are the absolute electric permittivities of the silicon region~$\Omega_{\text{Si}}$ and the vacuum region~$\Omega_{\text{v}}$ and $\vec{E}_{\text{r,Si}}$, and $\vec{E}_{\text{r,v}}$ are the electric fields in~$\Omega_{\text{Si}}$ and $\Omega_{\text{v}}$~\footnote{The electric field depends on the position~$\vec{r}$; thus, the maximum value in the denominator has to be intended as the maximum of the electric field density maxima in~$\Omega_{\text{Si}}$ and~$\Omega_{\text{v}}$.}. We obtain~$V_{\text{m}} \approx \SI{9.692e-17}{\meter\cubed}$. Hereafter, we refer to the integrals in the numerator of Eq.~(\ref{eq:Vm}) as~$I_{\text{Si}}$ and~$I_{\text{v}}$.

The zero-point electric field at a position~$\vec{r}$ can then be written as
\begin{equation}
\vec{E}_{\text{r}0} ( \vec{r} ) = \sqrt{\frac{h \tilde{f}_{\text{r}}}{2 \varepsilon_{\text{eff}} V_{\text{m}}}} \vec{G}_{\text{r}} ( \vec{r} ) ,
\end{equation}
where~$\vec{G}_{\text{r}} ( \vec{r} ) = \vec{E}_{\text{r}} ( \vec{r} ) / \max( \norm*{\vec{E}_{\text{r}} ( \vec{r} )} )$ is the electric mode function evaluated at~$\tilde{f}_{\text{r}}$ and $\varepsilon_{\text{eff}} = \varepsilon_{\text{Si}} q + \varepsilon_0 (1 - q)$ is the effective absolute electric permittivity; the filling factor~$q = I_{\text{Si}} / ( I_{\text{Si}} + I_{\text{v}} ) \approx 0.916$~\cite{Collin:2001}. For~$\varepsilon_{\text{Si}} / \varepsilon_0 = 11.9$, we obtain~$\varepsilon_{\text{eff}} / \varepsilon_0 \approx 10.981$.

Figure~\ref{Fig:App:Electric:field:simualtions}~(a) confirms that~$\norm*{\vec{E}_{\text{r}0}}$ is concentrated primarily around the resonator capacitor. Near the center region of the capacitor, we assume~$\vec{E}_{\text{r}0}$ to be uniform over the~$y$-axis and periodic over the~$x$-axis. As indicated in Fig.~\ref{Fig:App:Electric:field:simualtions}~(a), we consider an~$xz$ cross section intersecting three capacitor fingers. We discretize the cross section using a triangular mesh with triangles' side lengths~$\leqslant \SI{10}{\nano\meter}$; this mesh is much finer than in the rest of the simulated structure. This approach allows us to efficiently compute high resolution values of the electric field within the cross section,~$\vec{E}_{\text{r}0} (x,z)$.

We perform simulations for six different values of~$z$ with a spacing of~\SI{0.5}{\nano\meter}, from~$0.5$ to \SI{3.0}{\nano\meter} above the top surface of either the Al film or the Si substrate. The results for one value of~$z$ are shown in Fig.~\ref{Fig:App:Electric:field:simualtions}~(b). We generate~$g$ by evaluating~$\norm*{\vec{E}_{\text{r}0}}$ at randomly picked points~$(x,z)$ corresponding to Q-TLS positions.

\bibliography{bibliography}

\clearpage

\pagebreak

\begin{center}
\textbf{\large Supplemental Material for ``Fluctuation Spectroscopy of Two-Level Systems in Superconducting Resonators''}
\end{center}



\newcommand{\beginsupplement}{%
	\setcounter{section}{0}
	\renewcommand{\thesection}{S\arabic{section}}%
	\setcounter{subsection}{0}
	\renewcommand{\thesubsection}{S\Roman{subsection}}%
	\setcounter{subsubsection}{0}
	\renewcommand{\thesubsubsection}{S\Alph{subsubsection}}%
	\titleformat{\subsubsection}[block]{\bfseries\centering}{\thesubsubsection.}{1em}{}
	\setcounter{table}{0}
	\renewcommand{\thetable}{S\arabic{table}}%
	\setcounter{figure}{0}
	\renewcommand{\thefigure}{S\arabic{figure}}%
	\setcounter{equation}{0}
	\renewcommand{\theequation}{S\arabic{equation}}%
	}

\section*{S1: RESULTS FOR R2}
	\label{Sec:SM:RESULTS:FOR:R2}

\begin{figure*}[ht!]
	\centering
	\includegraphics{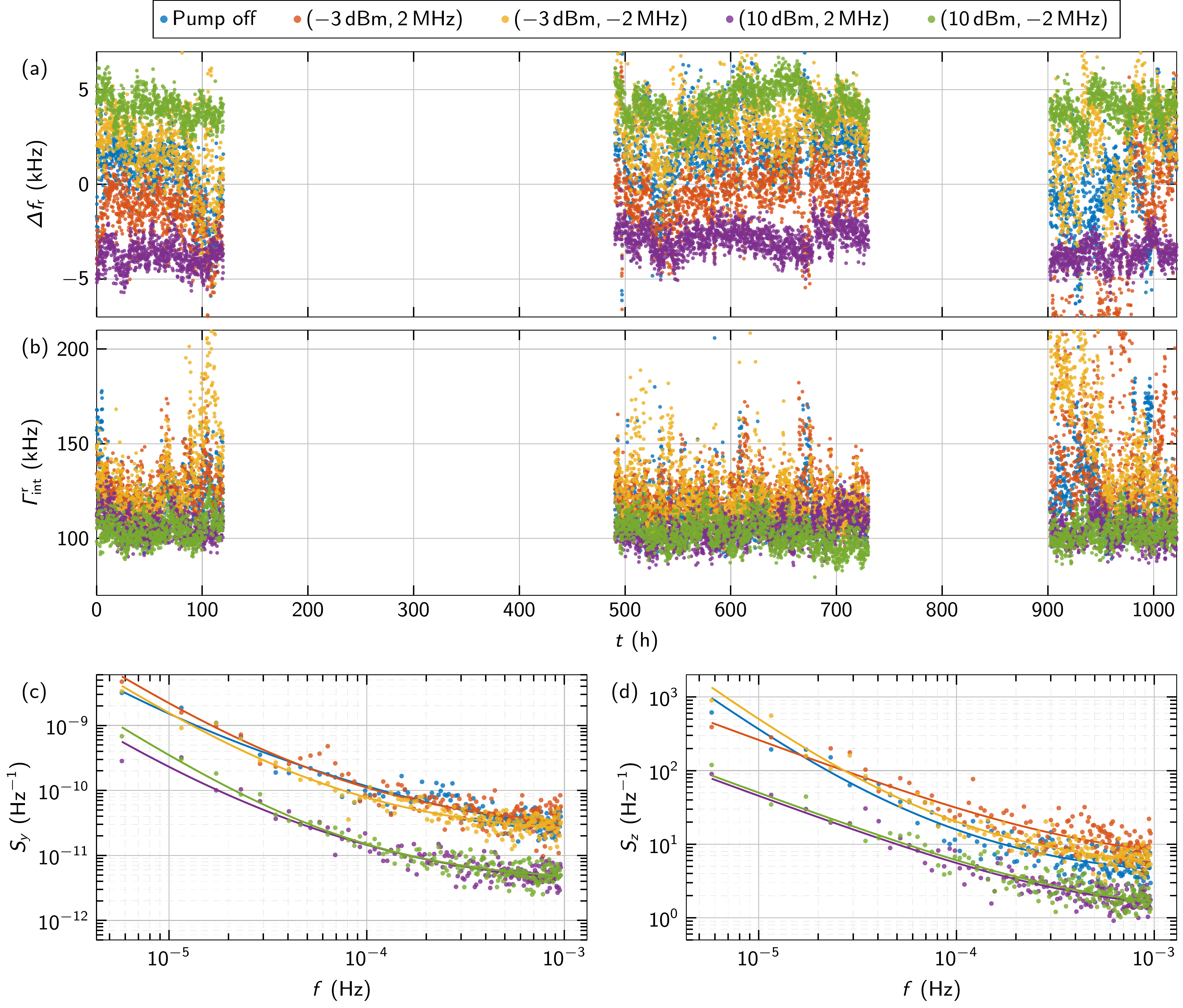}
		\caption{Fluctuation spectroscopy experiments for~R2. (a),(b) $\Delta f_{\text{r}}$ and $\Gamma^{\text{r}}_{\text{int}}$ vs.~$t$; the mean fit errors are~\SI{214}{\hertz} and \SI{2.7}{\kilo\hertz}, respectively. The four~$120$-\si{\hour} time periods are displayed in real time; the blank periods correspond to gaps between experiments. For presentation clarity, we clip the~$y$-axis, therefore loosing a few data points. (c),(d) $S_y$ and $S_z$ vs.~$f$. Dots correspond to PSD values and solid lines to fitting curves from Eq.~(8) in the main text. \label{Fig:SM:RESULTS:Fluctuation:spectroscopy:experiments:R2}}
\end{figure*}

Figure~\ref{Fig:SM:RESULTS:Fluctuation:spectroscopy:experiments:R2} shows the experimental time series~$\Delta f_{\text{r}} ( t )$ and $\Gamma^{\text{r}}_{\text{int}} ( t )$ as well as the PSDs of the normalized time series~$y ( t )$ and $z ( t )$, $S_y ( f )$ and $S_z ( f )$, for~R2.

\begin{figure*}[ht!]
	\centering
	\includegraphics{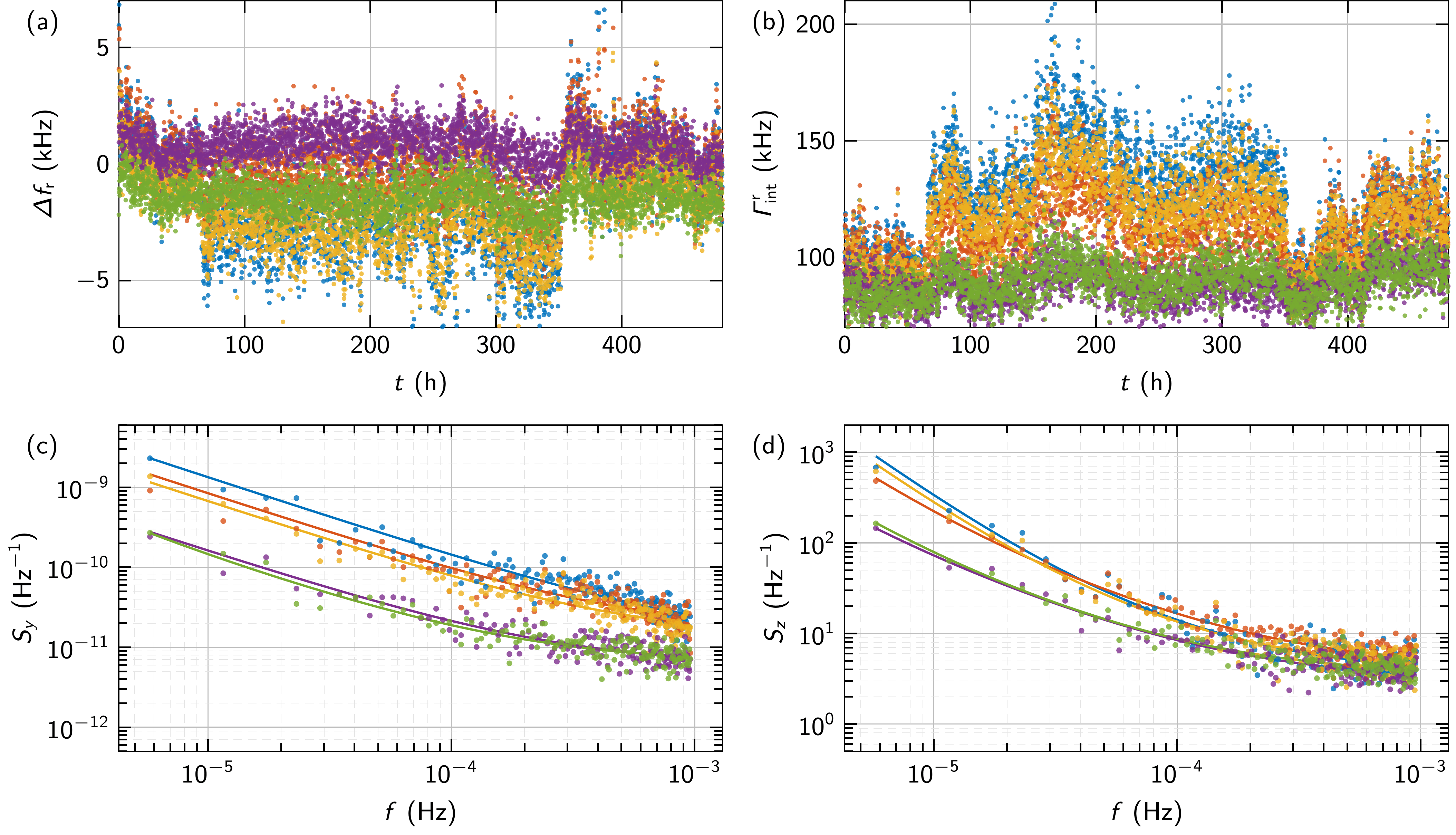}
	\caption{Fluctuation spectroscopy simulations for~R2. (a),(b) $\Delta f_{\text{r}}$ and $\Gamma^{\text{r}}_{\text{int}}$ vs.~$t$; we display only one~$120$-\si{\hour} time period. (c),(d) $S_y$ and $S_z$ vs.~$f$. Dots correspond to PSD values and solid lines to fitting curves from Eq.~(8) in the main text. The color coding is the same as in Fig.~\ref{Fig:SM:RESULTS:Fluctuation:spectroscopy:experiments:R2}. \label{Fig:SM:RESULTS:Fluctuation:spectroscopy:simulations:R2}}
\end{figure*}

Figure~\ref{Fig:SM:RESULTS:Fluctuation:spectroscopy:simulations:R2} shows simulations of the experiments reported in Fig.~\ref{Fig:SM:RESULTS:Fluctuation:spectroscopy:experiments:R2}.

\section*{S2: HIGH-POWER PUMP}
	\label{Sec:SM:HIGH-POWER:PUMP}

\begin{figure*}[ht!]
	\centering
	\includegraphics{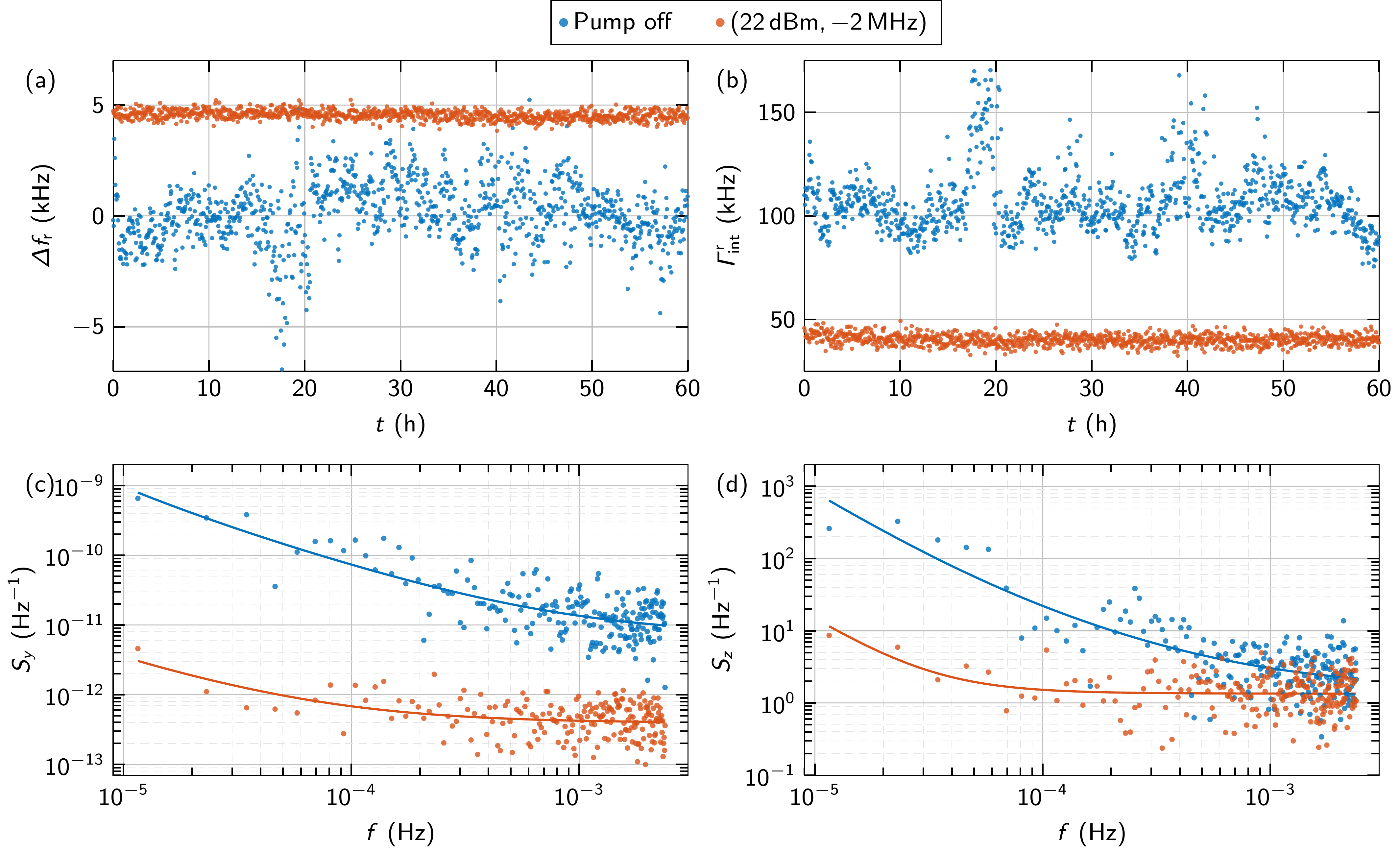}
		\caption{High-power pump for~R1 (different cooldown). (a),(b) $\Delta f_{\text{r}}$ and $\Gamma^{\text{r}}_{\text{int}}$ vs.~$t$. (c),(d) $S_y$ and $S_z$ vs.~$f$. Dots correspond to PSD values and solid lines to fitting curves from Eq.~(8) in the main text.
		\label{Fig:SM:HIGH-POWER:PUMP}}
\end{figure*}

Figure~\ref{Fig:SM:HIGH-POWER:PUMP} shows~$\Gamma^{\text{r}}_{\text{int}} ( t )$ and $f_{\text{r}} ( t )$ as well as~$S_y ( f )$ and $S_z ( f )$ for~R1 at two pump settings: pump off and $(\SI{22}{dBm} , \SI{-2}{\mega\hertz})$, that is, with a high-power pump. This experiment is performed during a different cooldown than for the experiments reported in the main text. In this cooldown, $\tilde{f}_{\text{r}} \approx \SI{5.556966}{\giga\hertz}$ and $\widetilde{\Gamma}^{\text{r}}_{\text{int}} \approx \SI{35}{\kilo\hertz}$. In these experiments, we measure only one~$60$-\si{\hour} time period; the PSDs are estimated using Welch's method with approximately~$24$-\si{\hour} segments, rectangular windows, and \SI{50}{\percent} overlap. The high-power results match well with our GTM-based simulations (not shown). These experiments indicate a noise reduction by more than two orders of magnitude, even for a detuning of~$\sim 100$ linewidths away from resonance.

\section*{S3: SETUP}
	\label{Sec:SM:SETUP}

\begin{figure}[t!]
	\centering
	\includegraphics{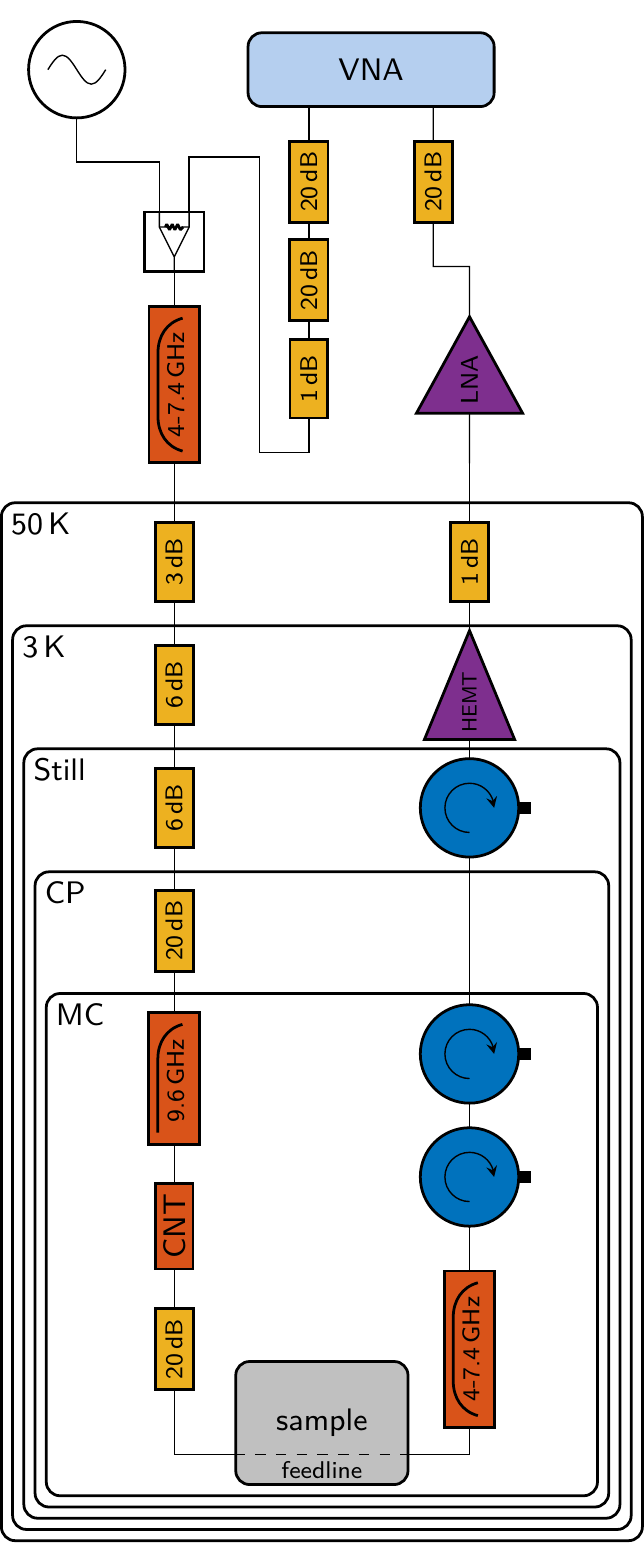}
		\caption{Setup schematic diagram. Yellow boxes indicate attenuators. Orange boxes indicate filters. CNT: carbon nanotube. Blue circles indicate circulators. Purple triangles indicate amplifiers. HEMT: high-electron-mobility transistor. LNA: low-noise amplifier. CP: cold plate. MC: mixing~chamber.
	\label{Fig:SM:Setup}}
\end{figure}

\begin{figure}[t!]
	\centering
	\includegraphics[width=\columnwidth]{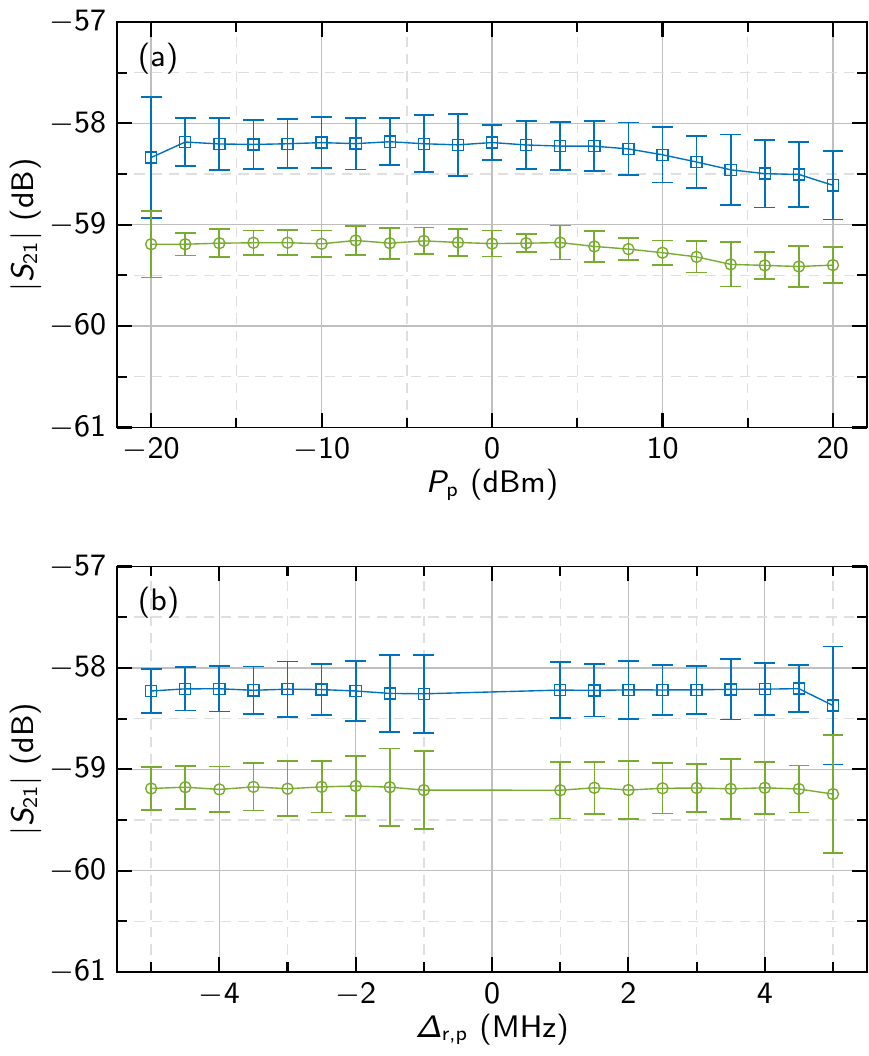}
		\caption{Two-tone compression for~R1 (blue squares) and R2 (green circles). (a) $\abs{S_{21}}$ vs.~$P_{\text{p}}$ at $\Delta_{\text{r,p}} = \SI{-2}{\mega\hertz}$. (b) $\abs{S_{21}}$ vs.~$\Delta_{\text{r,p}}$ at $P_{\text{p}} = \SI{5}{dBm}$; note that this second experiment is used to test for frequency nonuniformity in the amplification chain. Power and detuning ranges are as in Fig.~1 of the main text. Error bars represent~\SI{95}{\percent} confidence intervals. Lines are guides to the eye.
	\label{Fig:SM:Two-tone:compression}}
\end{figure}

Figure~\ref{Fig:SM:Setup} shows our experimental setup. The probe field power used in the experiments is always very low; however, the pump field power could reach large enough values to drive unwanted nonlinearities in the amplification chain (i.e., gain compression). Thus, we conduct \emph{two-tone compression} experiments to verify we are not saturating the amplifiers. These experiments are reported in Fig.~\ref{Fig:SM:Two-tone:compression}. Compression never exceeds~\SI{1}{\decibel} in any of our experiments.

\newpage

\newpage

\newpage

\newpage

\newpage

\end{document}